\documentstyle[11pt,fleqn]{article}
\newcommand{\N}{\mbox{$I\!\!N$}}             
\newcommand{\R}{\mbox{$I\!\!R$}}             
\newcommand{\C}{\mbox{$I\!\!\!\!C$}}         
\newcommand{\beq}{\begin{eqnarray}}          
\newcommand{\eeq}{\end{eqnarray}}          

\begin{document}


\hfill{\sl PREPRINT - UTM 558}
\par
\bigskip
\par
\rm


\par
\bigskip
\LARGE
\noindent
{\bf
 Proof of the symmetry of the off-diagonal Hadamard/ \ Seeley-deWitt's 
 coefficients in $C^{\infty}$ Lorentzian manifolds by a  ``local Wick 
 rotation''.}
\par
\bigskip
\par
\rm
\normalsize



\large
\smallskip

\noindent {\bf Valter Moretti}

\large
\smallskip

\noindent
Department of Mathematics, Trento University,
I-38050 Povo (TN), Italy.\\
E-mail: moretti@science.unitn.it


\par
\bigskip
\par
\hfill{\sl August 1999}
\par
\medskip
\par\rm



\noindent
{\bf Abstract:}
Completing the results achieved in a previous 
paper, we prove the symmetry of Hadamard/Seeley-deWitt  off-diagonal 
coefficients in smooth $D$-dimensional {\em Lorentzian} manifolds.
This result is relevant because it plays a central r\^{o}le in 
Physics, in particular in the theory of the stress-energy tensor 
renormalization procedure in quantum field theory in 
curved spacetime. 
To this end, it is shown that,  in any Lorentzian manifold, 
a sort of ``local Wick rotation'' of the metric can be performed  provided 
the metric is a (locally) analytic function of the coordinates and the 
coordinate are appropriate. No time-like Killing field is 
necessary. Such a local  Wick rotation analytically continues 
the Lorentzian metric in a neighborhood of any point (more generally,
in a neighborhood of a space-like (Cauchy) hypersurface)
into a Riemannian metric. The continuation locally 
 preserves geodesically convex 
neighborhoods. In order to make rigorous the procedure, the concept
of a complex pseudo-Riemannian 
(not Hermitian or K\"ahlerian) manifold is introduced
and some  features are analyzed. Using these tools,
the symmetry of  Hadamard/Seeley-deWitt  off-diagonal coefficients is proven
in Lorentzian analytical manifolds by analytical continuation
of the (symmetric) Riemannian heat-kernel coefficients. This continuation is
performed  in geodesically convex neighborhoods in common with both the 
metrics.
Then, the symmetry is generalized  to $C^\infty$ non analytic
Lorentzian manifolds by approximating Lorentzian $C^{\infty}$ metrics by
analytic metrics in common geodesically convex neighborhoods.
\par
 \rm





\section{Introduction, generalities and summary of previous results.}

{\bf 1.1.} In a previous paper  \cite{m3} we have considered the problem 
of the symmetry
of heat-kernel/Seeley-deWitt  coefficients, taken off-diagonal,
 for a second order 
differential  operator $A_0$ defined in a manifold ${\cal M}$.\\  
As is well-known \cite{wald78,wald94} that symmetry property 
assures the validity of some physically 
very important requirements (e.g. the conservation along the motion)
of the quantum stress-energy tensor in quantum field theory in curved spacetime,
whenever such a tensor is renormalized by means of the "point-splitting" procedure. In  \cite{m3},
 we considered the Euclidean case whereas, within this paper we want to 
deal with the Lorentzian case which is much more interesting on physical 
grounds.\\
From now on, ${\cal M}$  denotes a 
(real, Hausdorff, paracompact, connected,
 orientable) $D$-dimensional $C^{\infty}$ manifold endowed with a non-singular
 either Lorentzian (namely, the signature is
  $(-,+,\cdots,+)$) or Riemannian metric, 
${\bf g}$ \footnote{In the {\em gr-qc} 
version of \cite{m3}, 
we also assumed the positivity of $A_0$  in the Riemannian 
case and the geodesic completeness in general. Actually, these requirements 
are not necessary  to assure the symmetry of 
the heat-kernel coefficients and they can be dropped as can be shown with a 
little modification of {\bf Theorem 2.1} in \cite{m3}.}. 
(In the next section we shall consider also complex manifolds.)
The operator $A_0$ has the form
\begin{eqnarray}
A_0 = - \Delta + V\:\: : \:\: C_0^{\infty}({\cal M})  \rightarrow
L^2({\cal M}, d\mu_g) \label{caseR}\:,
\end{eqnarray}
whenever the metric is Riemannian.
Conversely, in the Lorentzian case, the operator $A_0$ has the form
\begin{eqnarray}
A_0 = - \Delta + V\:\: : \:\: D({\cal M})
\rightarrow C^{\infty}({\cal M})  \label{caseL}\:,
\end{eqnarray}
  $D({\cal M})$
being any domain of smooth 
functions, like $C^\infty_0({\cal M})$ or $C^\infty({\cal M})$.
$\Delta := \nabla_a\nabla^{a}$ denotes the Laplace-Beltrami operator and
 $\nabla$ means the covariant derivative associated to the metric
connection. $d \mu_{g} $ denotes the natural Borel measure induced by
the metric, and $V$ is a {\em real}
function of  $C^{\infty}({\cal M})$. (See \cite{m1,m2,m3} for discussions
concerning the existence and the relevance of self-adjoint extensions
of $A_0$ in $L^2({\cal M}, d\mu_g)$ in both cases.)
Throughout the text, if $({\cal U}, \vec{x})$ is a local chart of the 
differentiable structure of a 
$n$-dimensional manifold ${\cal M}$ and thus 
$\vec{x} : {\cal U} \to {\cal V} : p \mapsto 
(x^{1},\ldots,x^{n})(p) = \vec{x}(p)$,
${\cal V}\subset \R^n$, we shall indentify ${\cal U}$ with ${\cal V}$, 
writing $(x^{1},\ldots,x^{n})\in {\cal U}$ as well as
 $p\in {\cal V}$, whenever it does not give rise to misunderstandings.\\

\noindent 
The heat-kernel coefficients for the Riemannian case and the Seeley-deWitt
coefficients for the Lorentzian case, barring numerical factors,
coincide with the coefficients
which appear in the singular part of the Hadamard local solution (or Hadamard
parametrix) for the linear homogeneous 
equation  associated to the operator $A_0$ 
\cite{ch,ca,garabedian,fu,bd,wald94}
(see also \cite{m1,m2,m3} where the same notations used here are employed,
for further references and comments.)
 The heat-kernel/Seeley-deWitt coefficients are
 given by the following definition 
(See \cite{m3} for further comments and remarks and for the corresponding
differential recursive definition).\\

\noindent {\bf Definition 1.1.} {\em Within the  hypotheses on
${\cal M}$ and $A_0$ given above, in any  fixed open geodesically convex
neighborhood ${\cal N}\subset {\cal M}$, both the  heat-kernel 
(for the Riemannian case) and
 Seeley-deWitt (for the Lorentzian case)
 coefficients are 
the  functions defined on  ${\cal N} \times {\cal N}$ and
labeled
by $j\in \N$,
\begin{eqnarray}
a_{0}(x,y) &=&  \Delta_{VVM}^{1/2}(x,y)
\label{vvm}\:,\\
a_{(j+1)}(x,y) &=& - \Delta^{1/2}_{VVM} (x,y) \int_0^{1}
\lambda^{j}\left[  \Delta^{-1/2}_{VVM} A_{0x(\lambda)} a_j\right]
(x(\lambda), y) d\lambda\:.
\label{s}
\end{eqnarray}
$\lambda \mapsto  x(\lambda)$ is the unique
geodesic segment from $y \equiv x(0)$
to $x \equiv x(1)$ contained completely in ${\cal N}$.}\\

\noindent {\em Remark.} This definition can be given as it stands also in the 
general case of a {\em non-singular semi-Riemannian} metric, namely, 
when more than one eigenvalue of the metric is negative and no 
eigenvalue vanishes. This is a straightforward consequence of the theory 
developed in ${\bf 2.3}$ below.\\

\noindent  $\Delta_{VVM}(x,y)$ is a (smooth or analytic\footnote{Throughout 
this work "smooth" means $C^\infty$ and
"analytic" ($C^\omega$) means {\em holomorphic} whenever the considered 
functions are 
complex valued.}
 depending on the 
hypotheses on the metric)  {\em bi-scalar}
called the {\em van Vleck-Morette determinant} (see \cite{m3} for details). 
In any coordinate system $\vec{u}=(u^1,\cdots, u^D)$
defined in any open totally normal (or geodesically convex) neighborhood
${\cal N}$, if $x,y\in {\cal N}$ and $g := det g_{ab}$, we have
\begin{eqnarray}
\Delta_{VVM}(x,y) := \left[ (-1)^D\frac{g(\vec{x})}{|g(\vec{x})|}\right]
\: \frac{1}{\sqrt{g(\vec{x})g(\vec{y})}}
\det \left( \frac{\partial^2 \sigma(\vec{x},\vec{y})}{\partial 
x^{a} \partial y^{b}}
 \right) > 0
\label{vvmdef}\:.
\end{eqnarray}
Above $x\equiv \vec{x}$, $y\equiv \vec{y}$ and
$\sigma(\vec{x},\vec{y})$ 
is one half the "squared geodesical distance" of $x$ to $y$ (see \cite{m3} 
for details). The right-hand side of (\ref{vvmdef}) 
is positive  with the choice done for the first (constant) coefficient, 
not depending on the (fixed) {\em non-singular semi-Riemannian} signature of the metric 
(in particular, Riemannian or Lorentzian) and the used coordinates.\\

\noindent {\em Remarks.}\\ 
{\bf (1) }These definitions can be given also if, in any non-singular 
semi-Riemannian case, ${\cal M}$ denotes a manifold with (smooth) 
boundary $\partial 
{\cal M}$.   In this case it is also required that
 the  fixed open geodesically 
convex neighborhood ${\cal N}$ does not intersect $\partial {\cal M}$. 
The results obtained in this paper can be straightforwardly generalized
to manifolds with boundary.\\
{\bf (2)} Differently from {\bf Definition 1.1} in \cite{m3}, here we prefer
 to distinguish explicitly between the Lorentzian and the 
Riemannian case employing
a different nomenclature (heat-kernel or Seeley-deWitt coefficients 
respectively).\\
{\bf (3)} The coefficients defined by 
(\ref{s}) are either smooth if both the metric and $V$ are smooth or
 (real) analytic if both the metric and $V$ are (real) 
analytic (see \cite{m3}).\\

\noindent These coefficients have been 
showed to be symmetric in $x$ and $y$ whenever the metric and $V$ are
smooth (or analytic)
 and the metric is {\em Riemannian} \cite{m3}. This holds true despite the
non-symmetric definition (\ref{s}) and despite several subtleties in the 
convergence properties of the off-diagonal 
heat-kernel expansion which could be 
non-asymptotic. 
As we said previously, this result is physically relevant within the theory of the point-splitting 
renormalization of the stress-energy tensor in curved spacetime concerning
so-called Hadamard quasi-free quantum states. Indeed, the symmetry of the
heat-kernel or Seeley-deWitt coefficients trivially 
implies the symmetry of the coefficients which appear in the 
{\em singular part} of  (Euclidean or Lorentzian) Hadamard parametrix 
(see {\bf 1.3} of \cite{m3} for further details).
The symmetry property is a
sufficient\footnote{It is not so clear whether or not this condition is
necessary. After the appearance of the first version  of \cite{m3},
R.M. Wald pointed out to me that a weaker requirement
should be, in practice, sufficient (see comment before 
{\bf Proposition 2.1} in \cite{m3}).}
condition which assures a final well-behaved renormalized (Euclidean or
Lorentzian)
stress-energy tensor (see \cite{wald78,fsw} and references in \cite{wald94}).
Such important requirement has been assumed in the mathematical-physics 
literature without an explicit proof to the knowledge of the author
(see \cite{m3} for further comments).
This paper is devoted to show  that the symmetry holds true also in the 
Lorentzian case which is much more interesting on physical grounds. \\

\noindent {\bf 1.2.} {\em The rough idea of the proof of the symmetry for the
Lorentzian case.} In principle, a direct attempt to prove the symmetry
could be performed as we done in the Riemannian case \cite{m3}. 
That is, by employing the so-called Seeley-deWitt (or Schwinger-deWitt)
expansion of the integral kernel associated to the one-parameter 
group of unitary operators 
generated by some suitable self-adjoint extension of 
$A_0$ \cite{bd,fu,ca,m3}. In fact, the Seeley-deWitt coefficients are 
just  {\em the} coefficients
of this expansion. This expansion is the direct 
analogue of the heat-kernel expansion
 \cite{ch,m3}. Anyway, 
the convergence properties of the former are much
 more complicated than those of the latter (see discussions and references in
\cite{fu,m3}), so we prefer to 
follow an alternative way, which seems to be more interesting also on 
mathematical-physics grounds. The rough idea of the proof involves a sort of 
{\em "local" Wick rotation}, namely a somehow continuation of the relevant 
coefficients from the Lorentzian theory into the Riemannian one. 
By the uniqueness theorem of the analytical continuation, this should 
entail the generalization of the symmetry 
to the Lorentzian coefficients from the symmetry of Riemannian coefficients. 
We shall see that a sort of 
 "local" Wick rotation  can be performed, not depending on the presence of
time-like Killing vectors, provided the metric
and the potential $V$ are (real) analytic functions of the coordinates.
Finally, the generalization of the symmetry
 to the smooth {\em non-analytic}
 case can be obtained 
exactly as we done in the Riemannian case, making use of {\bf Proposition 2.1}
in \cite{m3}. This is, 
 by an approximation of smooth metrics by analytic metrics
and smooth functions $V$ by analytic functions. The intriguing issue
 is the generalization of the Wick rotation from Minkowski 
spacetime to curved {\em non-stationary} spacetimes. 
This is the argument of the next
section.

\section{"Local Wick rotation".}

{\bf 2.1.} {\em A generalized local Wick rotation.}
In QFT in flat spacetime the so-called (spatial) {\em Wick rotation} 
is a useful tool in spite of quite a vague definition. 
Roughly speaking, the Wick rotation is
 nothing but an analytical continuation of the
Minkowskian time coordinate into imaginary values: 
$t \to i \tau$ for $\tau \in \R$.
This is done in order to produce 
a Riemannian background where one can define the Euclidean QFT.
Formally, the metric changes as follows
\beq
\Phi_L = -dt\otimes dt + \sum_{i=1}^dd x^{i} \otimes d x^{i} 
\:\:\: \to\:\:\: \Phi_R = d\tau \otimes d\tau + \sum_{i=1}^d
 dx^{i} \otimes dx^{i} 
\label{wick} \:.
\eeq
Such a procedure is performed for several goals, e.g., to make sensible 
the path integral as a Wiener measure or to build up the thermal QFT.
In {\em curved} spacetime, 
the use of the Wick rotation is much more problematic.
In particular, there is no guarantee 
for the fact that the continued Riemannian 
metric is {\em real} whenever the initial {\em Lorentzian} metric is real.
In principle, a somehow sufficient condition which gives a real Riemannian 
  metric is given by the 
requirement of a {\em static metric} \cite{wald79,FR,fu,wald94}. 
If the metric is only {\em stationary} the Wick rotation is more problematic
 and generally involves the analytic
continuation of further parameters than the time coordinate \cite{ha}.
In spite of these difficulties, the Wick rotation is successfully 
 used in QFT in curved spacetime, Quantum Gravity and black holes theory, 
where it is a very powerful tool in studying  black hole 
thermodynamics in particular \cite{EQG}. \\
In this work, to get the proof of the symmetry of Seely-deWitt coefficients,
 we want to generalize the Minkowskian Wick rotation
to quite a  general Lorentzian manifold dropping any hypothesis concerning
the presence of temporal Killing fields. As far as
we are concerned, a sort of  Wick rotation of the {\em metric} 
is sufficient to prove the first step of our symmetry theorem. 
To this end, let us focus our attention on (\ref{wick})  once again. 
Notice that the same result can be obtained
by an analytic continuation of the {\em metric} rather than the time 
coordinate. In fact, we are free to interpret (\ref{wick}) as
\beq
\Phi_L = g_{Lab}dx^{a} \otimes dx^{b}
 \:\:\: \to \:\:\: \Phi_R = g_{Rab}dx^{a}\otimes dx^{b}
\label{vick2} \:.
\eeq
where 
\beq
g_{Lab} = diag(-1,1,1,1)\:\:\:\: \mbox{and} \:\:\:\:
g_{Mab} = diag(1,1,1,1)\:.\nonumber
\eeq
Differently from the customary interpretation, the metric has now changed, 
since the eigenvalue $-1$ has been continued into
a final eigenvalue $+1$, but the manifold has 
remained the initial one. The changes have
taken place only in each {\em (co-)tangent fiber}.
In {\em this} sense the continuation is "local".
In principle, such a procedure could be used also in curved spacetime
without the requirement of a static or stationary metric.
We are not interested in the issue about which properties of the 
customary interpretation are preserved by the new interpretation. 
Only two important points have to be remarked: Following the new 
procedure, there is no guarantee for the fact that the  Riemannian metric so
obtained is a vacuum solution of Euclidean Einstein  equation if the initial Lorentzian metric
 is a vacuum solution of the Lorentzian 
Einstein equations; moreover it is worthwhile stressing that the 
found procedure is very non-unique. \\
The use of complex metrics is compulsory if one assume the
absence
of pathologies in the structure of geodesically convex neighborhoods 
during the 
continuation procedure. Such an absence is essential as far as our
main goal, i.e., the proof of symmetry of Hadamard coefficients, is concerned,
because these coefficients are defined just 
in geodesically convex neighborhoods.
 If we want to  pass from a
 Lorentzian to a Riemannian metric continuously, the signature has to
change from $(-,+,\cdots,+)$ to $(+,+,\cdots,+)$. Therefore, the determinant
of the metric, in any fixed coordinate frame, must change sign somewhere. 
This implies that, employing only {\em real} metrics, 
some of these must become {\em singular} 
somewhere during the continuation. Therefore,
pathologies would arise concerning the exponential maps and the structure 
of geodesical neighborhoods. On the other hand, the use of {\em complex} 
coefficients of the metric makes sense only in complex manifolds.
For instance, in general, the equations of the geodesics admit no solution 
in {\em real} coordinates with
{\em complex} coefficients of the metric. Therefore, we are forced to 
extend the initial Lorentzian metric and manifold to complex values 
in complex coordinates. A natural
way to do this follows from the assumption of an initial {\em real analytic}
metric. The "complex metrics" which arise in the continuation procedure
are not Hermitian or K\"ahlerian but generalize the concept of a 
pseudo-Riemannian metric into a complex context.\\

To proceed with our idea of a local generalized Wick rotation
we need a well-known preliminary definition.\\

\noindent {\bf Definition 2.1.}
{\em Let $({\cal M}, {\bf g})$ be a  $C^{k}$ ($k\in 
\{2,\ldots,\infty,\omega\}$) Lorentzian 
$(D=d+1)$-dimensional manifold.  Choose any embedded 
spacelike hypersurface ${\cal S}$ and an open  neighborhood 
${\cal O}$ such that ${\cal O}\cap {\cal S} \neq \emptyset$.
Taking ${\cal O}$ sufficiently small (preserving the condition above) 
if necessary, any admissible ($C^k$) 
local coordinate system defined in ${\cal O}$,
 $\vec{x} = (x^{0},\ldots,x^d)$ with, $(x^{1},\ldots,x^d) \in \Omega$
 open subset of $\R^{D}$ and $x^{0}\in \:]-\delta,\delta[\:$, $\delta>0$,
 such that, in ${\cal O}$,
\begin{eqnarray}
{\cal S}\cap {\cal O} &=& \{ (x^{0},\ldots,x^n)\in 
\:]-\delta,\delta[\:\times \Omega \:\:|\:\: 
x^{0} = 0\} \label{Fermi0}\:,\\
g_{00}&=& -1 \label{Fermi1}\:,\\
g_{0a}&=&0 \:\:\:\:\mbox{for}\:\:\: a>0 \label{Fermi2}\:.
\end{eqnarray}
 is said 
{\bf local synchronous coordinate system with respect to ${\cal S}$}.}\\

\noindent Concerning the existence of such
 coordinate systems, see \cite{waldR} and, for a more general mathematical 
discussion, see Chap.7 of \cite{oneill} where these coordinates 
are called "Fermi coordinates with respect to a given spacelike 
hypersurface". A sketch of the proof of their existence will also 
be given within the proof of  {\bf Theorem 2.2}.\\

\noindent {\em  Remark.} Any point $p\in {\cal M}$ belongs to
 an embedded spacelike hypersurface: Such a hypersuface can be obtained as
  ${\cal S}_{p} = \{ exp_{p}(X^{a} {\bf e}_{a})\:\:|\:\: X^{0}=0\:
  \:(X^{1},\ldots,X^d)\in \Omega\}$, where $\Omega$ is 
   a suitable small neighborhood 
 of the origin of $\R^d$,
 $({\bf e}_{0},\ldots, {\bf e}_{d})$ being an orthonormal base of 
$T_{p}({\cal M})$ with $({\bf e}_{0},{\bf e}_{0}) = -1$.\\
Notice that the found hypersurface about $p$ is not uniquely defined. \\

\noindent  Let us consider an analytic manifold ${\cal M}$ endowed
with an analytic Lorentzian metric ${\bf g}$. Take a complex 
analytic continuation of a synchronous coordinate system 
defined in a open neighborhood ${\cal O}$ 
about a point $p\in {\cal M}$, $ \vec{z}= 
(z^0,z^1,\cdots, z^d)$ 
with $z^{a}= x^{a} + i y^{a} $ into a complex open neighborhood ${\cal 
G} \in 
\C^D$ (containing the initial real domain of definition of the coordinates).
 Suppose that  analytic continuations of the
 the functions  $\vec{z} \mapsto g_{ab}(\vec{z})$ are defined in 
 ${\cal G}$. In general 
$\vec{z}\mapsto g_{ab}(\vec{z})$ are complex-valued  functions
but preserve (\ref{Fermi1}) and (\ref{Fermi2}). 
By these functions it is possible to define  a {\em non-singular}
 "complex pseudo-metric" (the rigorous definition will be given in 
{\bf Definition 2.2})
 $\vec{z}\mapsto {\bf g}(\vec{z}) = g_{ab}(\vec{z}) \:dz^a\otimes dz^b$
 on ${\cal G}$. Finally, fix an arbitrary 
 real $\lambda >0$ and consider the class of "complex
pseudo-metrics"  
$\left\{ {\bf g}_{(\lambda\theta)}\right\}_{\theta}$ 
where $\theta \in \C$, defined in the coordinates of ${\cal G}$ by 
\beq
g_{(\lambda \theta) 00}(\vec{z}) &:=&  g_{00}(\vec{z}) 
\lambda^{2\theta/\pi} e^{i\theta}\:, \label{analytic1} \\
g_{(\lambda \theta) ab}(\vec{z}) &:=&    g_{ab}(\vec{z})\:\:\:\: \mbox{for} \:\:\:\:(a,b)
\neq (0,0) \label{analytic2}\:.
\eeq
We want to use  this class to continue the initial {\em Lorentzian} 
metric obtained for $\theta = 0$,
${\bf g}
 = {\bf g}_{(\lambda 0 )}$ into  a final {\em Riemannian}
 "Wick-rotated metric", obtained for 
$\theta = \pi$.
Indeed, within our hypotheses 
 the "Wick-rotated metric"
$\bar{\bf g}_{\lambda}(\vec{z}):= {\bf g}_{(\lambda\pi)}(\vec{z})$ defines 
a {\em real}, {\em non-singular} and {\em Riemannian} metric
 for $\vec{z} = \vec{x}\in {\cal O}$ when it acts on {\em real}
 (with respect to the considered coordinates) vectors.
In particular, fixed any positive real $\lambda$,
 ${\bf g}_{(\lambda \theta)}(\vec{z})$ is {\em non-singular} in a 
 complex open neighborhood of
 $[0, \pi]\times {\cal O}$. More strongly,
 in a sense we shall specify later, fixed the parameter $\lambda$,
the procedure {\em preserves geodesically convex 
neighborhoods} for complex value of $\theta$. 
 In practice, the presented procedure locally 
defines an {\em analytic} continuation of metrics\footnote{Notice
that the functions $(\theta,\vec{z})\mapsto g_{(\lambda 
\theta)ab}(\vec{z})$ are indeed  
{\em analytic} in $(\theta,z)$.} which interpolates through complex metrics,
 from {\em Lorentzian} to {\em Riemannian}
metrics and preserves the  local geodesical structure at each step.
 As we shall see, the apparently superfluous
 parameter $\lambda$  plays a central r\^{o}le in using the local 
  Wick rotation to get the symmetry of Seeley-deWitt coefficients.\\

\noindent Let us state some of the results argued above into a precise 
theorem.\\

\noindent {\bf Theorem 2.1.} {\em Let $({\cal M}, {\bf g})$ be 
a $(D=d+1)$-dimensional Lorentzian manifold with class
$C^\omega$. Let  ${\cal O}\subset {\cal M}$ be any open set 
endowed with local synchronous  coordinates (with respect to some 
embedded hypersurface) $\vec{x} = (x^0,\cdots,x^d)$ and consider the 
coefficients of the metric
$g_{ab}(\vec{x})$ in these coordinates. Fix a real $\lambda>0$.\\ 
 Then, there is a complex open set ${\cal G} \subset \C^D$ 
endowed with a differentiable structure induced by 
coordinates $\vec{z} = (z^0,\cdots,z^d)$ with $z^{a} = x^{a} + i y^{a}$, 
$a=0,\cdots,d$ and ${\cal O} \subset {\cal G}$ (in the obvious sense), 
where the components 
of the metric $\vec{x}\mapsto g_{ab}(\vec{x})$, $a,b = 0,\cdots,d$,
can be analytically continued 
into analytic complex functions $\vec{z}\mapsto g_{ab}(\vec{z})$. 
  Moreover  the functions defined in $\C \times {\cal G}$, 
$\vec{z}\mapsto g_{(\lambda \theta)ab}(\vec{z})$,
where $g_{(\lambda \theta)ab}(\vec{z})$ have been defined in} 
(\ref{analytic1}) {\em and} (\ref{analytic2}),
 {\em define a $\theta$-parametrized  class 
$\{ {\bf g}_{(\lambda \theta)}\}_{\theta}$, $\theta\in \C$, 
of complex analytic $(0,2)$-degree symmetric fields
\beq
{\bf g}_{(\lambda \theta)}(z) := 
 g_{(\lambda \theta)ab}(\vec{z})\:dz^a\otimes dz^b\:, \label{otimes}
\eeq
($z\equiv \vec{z}$) which are non-degenerate 
everywhere in  the complex manifold ${\cal G}$.\\
In ${\cal O}$, this class analytically continues in the parameter $\theta,$ the 
initial Lorentzian metric ${\bf g}= {\bf g}_{(\lambda 0)}$
into the Riemannian {\bf Wick-rotated  metric}}
 \begin{eqnarray}
 \bar{\bf g}_\lambda := {\bf g}_{(\lambda\pi)} \label{gbar}\:.
 \end{eqnarray}\\

\noindent {\em Proof.} It is straightforward if one takes into account 
that the components of $g_{(\lambda \theta)ab}(\vec{z})$ with $a,b>0$ do not 
depend on $\theta$ and $\lambda$ and also noticing that
 (\ref{Fermi1}) and (\ref{Fermi2}) hold true for the "complex-continued metrics"
   in (\ref{otimes}). Hence,
$$|\det\{[g_{(\lambda \theta)ab}(\vec{z})]_{a,b= 0,\ldots,d}\}|
= |e^{\theta\:[i + (2/\pi) \ln \lambda ]}|\:|\det\{[g_{ab}(\vec{z})]_{a,b= 
1,\ldots,d}\}|\:.$$ The former factor on the right-hand side is 
positive and cannot 
vanish if $\lambda>0$ whatever $\theta \in \C$ and the latter factor
defines a  continuous function of the only variable 
$\vec{z}\in {\cal G}$ which preserves the positive sign in a 
open complex neighborhood of each point $\vec{z}=\vec{x} \in {\cal O}$
where the function  is positive by hypotheses.
We can redefine ${\cal G}$ as the union of all of these open 
neighborhoods.    $\:\:\Box$\\
 
\noindent {\em Remark.} Notice that we have used a little misuse of notations
in the last statement of the theorem.
Indeed, the initial Lorentzian metric ${\bf g}(x) = g_{ab}(\vec{x})
\:dx^a\otimes dx^b$ 
and the Wick-rotated 
Riemannian metric $\bar{\bf g}_\lambda(x) = \bar{g}_{(\lambda)ab}(\vec{x}) 
\:dx^a\otimes dx^b$ 
are defined in the {\em real}, with respect to the base induced by the 
coordinates specified above $z^a = x^a+iy^a$, cotangent space of ${\cal 
O}$. 
Conversely, all interpolating fields (including those
corresponding to the values $\theta=0,\pi$) 
${\bf g}_{(\lambda \theta)}$ in (\ref{otimes})
are defined in the whole {\em complex}
cotangent space of ${\cal O}$. Anyway, we shall use these notations also in the 
following since it does not produces misunderstandings.\\

\noindent{\em Note.} The local Wick rotation we have defined above
 can be generalized to  
a more large class of coordinates with a 
precise physical meaning, namely, coordinates where $x^{0}$ represents 
a "true" time and $x^{1},\ldots,x^d$ represent "true" spatial 
coordinates. In other words, in these coordinates,
it must hold $g_{00}<0$ as well as $g^{00}<0$.
Within this general approach, fixing a point 
$p\in {\cal M}$, the parameter $\lambda$ is related to the relative 
velocity between the rest reference $d$-dimensional space (subspace of 
$T_{p}({\cal M})$) of the 
infinitesimal observer evolving along $\partial_{x^{0}}|_{p}$ and the 
$d$-dimensional reference space (subspace of 
$T_{p}({\cal M})$) "normal" to the vector $dx^{0}|_{p}$.
Also {\bf Theorem 2.1} and {\bf Theorem 2.4} below can be generalized 
for these "physical" 
coordinates but the proofs are much more complicate. 
\footnote{A general discussion on these arguments, and all the relevant proofs,
can be found in the first {\em gr-qc} version of this paper.}.
We  also have the following almost straightforward result, 
which is interesting by its own not depending on our final goal.
It shows that the local procedure defined above can be relatively 
globalized about any spacelike embedded hypersurface (e.g. a Cauchy
surface if ${\cal M}$ is globally hyperbolic.) also when {\em global}
synchronous coordinates with respect to it do not exist. 
Anyway, we stress that we shall use the local result only in the proof 
of symmetry theorem.\\

\noindent {\bf Theorem 2.2.} {\em Let $({\cal M},{\bf g})$ be a
$(D=d+1)$-dimensional Lorentzian manifold with class $C^\omega$
and time-oriented. Fix any real $\lambda > 0 $.
Let  ${\cal S}\subset {\cal M}$ any fixed embedded space-like hypersurface 
with class $C^\omega$.\\ 
Then, there is an open 
$D$-dimensional open Lorentzian sub-manifold of ${\cal M}$, 
${\cal N}$ containing ${\cal S}$ which
admits the class ${\bf {\cal A}}({\cal N})$  
of ($C^\omega$) 
{\em time-oriented local synchronous coordinates with respect to $\cal S$} 
as an atlas. 
Moreover, the Wick-rotated metrics defined by} (\ref{gbar})
{\em in each local coordinate system of ${\bf {\cal A}}({\cal N})$ 
induce a {\em global} Riemannian $C^\omega$ metric on the whole 
sub-manifold ${\cal N}$.}\\

\noindent {\em Sketch of Proof.} See the {\bf Appendix} $ \:\:\Box$\\

In the next part we shall consider some features of the Wick-rotated 
metrics. To this end we need some definitions and results concerning 
complex metrics in complex analytical manifolds.\\

\noindent {\bf 2.3.} {\em Complex pseudo-Riemannian manifolds.}
To give a precise status to the complex field ${\bf g}_{(\lambda \theta)}$
defined on the complex manifold ${\cal G}$ presented in {\bf Theorem 2.1},
we introduce the concept of a {\em complex pseudo-Riemannian manifold}
and a {\em complex pseudo-Riemannian metric}.
Also with different nomenclature, several results obtained in the 
following can be found in \cite{lb} (see also \cite{cassa}).
First of all, we give some results concerning the existence and the 
analyticity of the exponential map in complex manifolds with a 
generally complex pseudo-Riemannian 
metric. Afterwards we discuss the existence of geodesically convex
neighborhoods and related features.  Let us start  
by giving the definition of a complex pseudo-Riemannian manifold.\\

\noindent {\bf Definition 2.2.}
{\em A complex analytic manifold ${\cal M}$} (\cite{kn}) 
{\em endowed with an analytic non-degenerate $(0,2)$-degree symmetric
 tensorial field ${\bf g}$ is said {\bf complex pseudo-Riemannian manifold} 
and the field ${\bf g}$ is said {\bf complex pseudo-metric}.
The complex pseudo-metric induces a {\em non-degenerate} complex quadratic 
form ${\bf V}\mapsto {\bf g}(z)({\bf V},{\bf V})$, 
 in the  tangent space  $T_{z}({\cal M})$ at any point $z\in {\cal M}$.
  We call such a quadratic 
form the {\bf complex pseudo scalar product} induced in $T_{z}({\cal M})$
by the complex pseudo-metric}.\\ 
 
\noindent {\em Remarks.} \\
{\bf (1)} It is worthwhile stressing 
that the complex pseudo scalar 
product induced on the tangent spaces is {\em not} 
Hermitian and the metric is {\em not}  K\"ahlerian. \\
{\bf (2)} It is clear that the manifold ${\cal G}$ introduced in {\bf Theorem 
2.1}, endowed with
any fixed field ${\bf g}_{(\lambda \theta)}$, is a complex pseudo-Riemannian
manifold.\\
{\bf (3)} The equations of the geodesics take the usual formula with the 
difference that the connection coefficients of the 
Levi-Civita connection (see below) induced by the complex 
pseudo-metric are complex analytic functions of the considered
coordinates.\\

\noindent Concerning the equation of the geodesics we can apply 
the following general lemma.\\

\noindent {\bf Lemma 2.1.} {\em Let $f : (z,Y,\alpha) \mapsto 
f(z,Y,\alpha) \in \C^{n}$ be a function in  $C^{\omega}(\bar{{\cal C}}; 
\C^{n})$ \footnote{Notice that $\bar{\cal C}$ is closed. We say that $f$ is 
analytic in a {\em closed} set  when  it is possible to continue $f$ into 
an analytic function defined in a 
{\em open} set which includes the  closed set.},
with ${\cal C} = B_{r_{1}}(z_{0})\times B_{r_{2}}(y_{0})  \times 
B_{r_{3}}(\alpha_{0})$ 
 where $B_{r_{1}}(z_{0}),  B_{r_{3}}(\alpha_{0}) \subset \C$
  and $ B_{r_{2}}(y_{0}) \subset \C^{n}$  
 are open balls with radii $r_{1},r_{2},r_{3} >0$ centered in 
 $z_{0},y_{0}, \alpha_{0}$ respectively.
 Consider the differential equation system depending on the
 parameter $\alpha \in \bar{B}_{r_3}(\alpha_0)$
 \begin{eqnarray}
 \frac{dY}{dz} =f(z,Y,\alpha) \:\:\:\:\:\:\:\:\:\:\:
Y\in C^{1}(B_{r'_{1}}(z_{0});\C^n) \:\:\:\:\:\: \mbox{for some }\:\:\:\
r'_{1}>0, \:\:r'_{1}< r_{1} 
  \label{equazione}
 \eeq
and  initial condition
\begin{eqnarray}
Y(t_0) = \bar{y}_0,\:\:\:\:\:\:\: \bar{y}_0 \in \bar{B}_{r'_{2}}(y_0),
 \:\: \:\: \mbox{where}\:\: r'_{2}>0, \:\: \:\:\mbox{is fixed and }\:\:\:\:
r'_{2}< r_{2} \label{condizione}\:.
\end{eqnarray}}

(a) {\em A solution of Eq.} (\ref{equazione}) {\em with initial
condition}
(\ref{condizione}) {\em exists and is unique in any set
$\bar{B}_{r'_{1}}(z_{0})$, provided that
\begin{eqnarray}
0<r'_{1} < Min\left(r_{1},\delta',\delta''\right) \label{delta}\:,
\end{eqnarray}
where 
\beq
\delta' = (r_{2}-r'_{2})/
Sup\left\{ ||f(z,y,\alpha)||\:\:|\:\:
(z, y, \alpha) \in \bar{{\cal C}} \right\} \nonumber\:,
\eeq 
 and
\beq
\delta'' = 1/Sup\left\{2 \sqrt{n\: Tr  \nabla {f^*}(t,y,\alpha)^{T}
 \nabla f(t,y,\alpha)}\:\:|\:\:(z,y,\alpha)\in \bar{\cal C} \right\}
\nonumber\:.
\eeq}

(b) {\em This solution
  satisfies $Y(z) \in \bar{B}_{r_{2}}(y_0)$ 
for any
$z \in  \bar{B}_{r'_{1}}(z_{0})$, whatever
$\bar{y}_0\in \bar{B}_{r'_{2}}(y_0)$ and
$\alpha \in \bar{B}_{r_{3}}(\alpha_{0})$.}

(c) {\em Moreover, varying also $y_0$ and $\alpha$, and writing down the 
dependence on these variables explicitly,
the function $(t,\bar{y}_0,\alpha)\mapsto Y(t,\bar{y}_0,\alpha)$
is analytic. In particular, it
belongs to $C^{\omega}(\bar{B}_{r'_{1}}(z_{0})\times \bar{B}_{r'_{2}}(y_{0})
  \times  \bar{B}_{r'_{3}}(\alpha_{0}))$ for any $r'_3>0$ with $r'_3<r_3$.}\\
 
\noindent {\em Proof.} See the {\bf Appendix}.\\

\noindent We can apply the lemma above to the equations of the geodesics
for a complex pseudo-metric ${\bf g}$ in coordinates $\vec{z} = 
(z^{1},\cdots, z^{D})$. For the moment we do not consider the 
further parameter $\alpha$.
The {\em first-order} geodesical equation system reads,
for the complex pseudo-metric ${\bf g}$ in the coordinates $\vec{z} = 
(z^{1 },\cdots,z^{D})$,
\begin{eqnarray}
\frac{d z^{a}(t,\vec{y},{\bf V})}{dt} &=&
U^{a}(t,\vec{y},{\bf V}) \label{geo1} \\
\frac{d U^{a}(t,\vec{y},{\bf V})}{dt} &=& -\Gamma_{bc}^{a}(\vec{y})
 U^{b}(t,\vec{y},{\bf V})  U^{c}(t,\vec{y},{\bf V})
\label{geo2}\:,
\end{eqnarray}
for $a=1,\cdots,D$.
(The sum over the
repeated indices is understood).
Above, the complex Levi-Civita connection coefficients are defined, as 
usual, by
\beq
\Gamma_{bc}^{a}(\vec{z}) := \frac{1}{2}g^{ad}(\vec{z})
\left( \frac{\partial g_{db}(\vec{z})}{\partial z^c} 
+ \frac{\partial g_{cd}(\vec{z})}{\partial z^b}  - \frac{\partial 
g_{bc}(\vec{z})}{\partial z^d}  
\right) \label{connection}\:,
\eeq
$\vec{y}$ and ${\bf V}$
 are, respectively, the initial position and the
initial velocity of the geodesic segment 
evaluated at $t=0$. The equations (\ref{geo1}) and (\ref{geo2})
 for $t\in \C$, locally
 admit a unique solution which satisfies the given initial 
conditions. 
 The existence  of sets where the solution exists, is unique and is 
 analytic  is assured by {\bf Lemma 2.1}.
 Let us indicate 
the local solution of the system above by
\beq
t \mapsto \gamma(t,\vec{z},{\bf V}) \:,\label{geodesic}
\eeq
for $t\in B_{\delta}(0)$,
 $\delta>0$, $(\vec{y},{\bf V})\in B_{\rho}(\vec{y}_{0})\times B_r({\bf 0})$, 
$\rho, r>0$.
Exactly as in the real Riemannian case, 
for any fixed complex number $c\neq 0$,
(\ref{geo1}) and (\ref{geo2}) entail the identity
\begin{eqnarray}
\gamma( c t,\vec{z},{\bf V}/c)
= \gamma(t,\vec{z},{\bf V})  \label{scaling}\:.
\end{eqnarray}
This means that if, for instance, $2 > \delta >0$, 
passing to the new variable $t'
= (2/\delta) t$, we can work with geodesics
defined in the interval $t'\in B_{2}(\vec{0})$ provided $r$ is replaced by
$r' = (\delta/2) r < r$. This can be done  preserving all remaining properties
concerning the analyticity and
 not depending on the initial condition  $\vec{z}$
(and not depending on any further parameter $\alpha$ as that in {\bf Lemma
2.1}).
 Since there is no ambiguity we can  use the name
$r$ instead of  $r'$ and $t$ instead of $t'$.
Therefore, from now on,   we suppose
$t \in B_{2}(\vec{0})$.
With this choice, a maps (\ref{geodesic}) with the restriction $t\in
\bar{B}_{1}(\vec{0})$ will be called {\em complex geodesic segment}.
It is worthwhile stressing that this is not a "usual" segment
because the parameter $t$ corresponds to {\em two} real parameters.\\
A complex geodesic segment  restricted to the real axis in the domain,
$s \in \R$, 
\beq
s \mapsto \gamma(s,\vec{z},{\bf V}) \:,   \:\:\:\: \mbox{where } \:\: 
s \in [0,1]\label{geodesicl}
\eeq
will be called {\em real-parameter geodesic segment}. Notice that it satisfies 
(\ref{geo1}) and (\ref{geo2}) in the variable $s$ and is real 
analytic in this variable. It determines the whole complex geodesical
segment by analytic continuation. Obviously, these definitions does not depend
on the chosen coordinates, so sometimes 
we shall use, e.g., $p$ instead of $\vec{y}$ in the second argument of a 
geodesic segment.\\
The {\em exponential map} is, as usual,  given as the analytic  map, defined in
an opportune  {\em open} set 
\beq
E = \bigcup_{p\in {\cal M}}
 \{p \}\times E_{p}
\subset T({\cal M}) \label{E}\:,
\eeq
where $E_{p}$ is an open neighborhood of the origin of 
 $T_{p}({\cal M})$ we shall specify shortly,
\beq
exp : E  \to {\cal M} : (p,{\bf V})
\mapsto \gamma(1, p, {\bf V}) \label{exp0}\:.
\eeq
The exponential map {\em centered in $ p\in {\cal M}$} is the map 
\beq
exp_{p} : E_p
 \to {\cal M} : {\bf V}
\mapsto \gamma(1,p, {\bf V}) \label{exp}\:.
\eeq
Obviously, these definitions does not depend on the used coordinates 
and, changing the domains one finds restrictions or 
extensions of the same function. Since, for any point $p 
\in {\cal M}$, it holds
\beq
 d (exp_{p})_{\bf 0}{\bf V} = \frac{d}{dt}|_{t=0} exp_{p}(t {\bf V})  =
 \frac{d}{dt}|_{t=0} \gamma(1, p, t{\bf V}) = \frac{d}{dt}|_{t=0} 
 \gamma(t, p, {\bf V})  = {\bf V} \nonumber\:,
 \eeq
there is a  open neighborhood, which we can assume to be {\em starshaped},
of the origin of the tangent space at 
$p$, where 
the exponential map centered in $p$ defines an analytic diffeomorphism onto a 
neighborhood of $p$ in the manifold. 
With an open starshaped neighborhood of ${\bf 0}$  
we mean an open neighborhood of ${\bf 0}$ such 
that if ${\bf V}$ belongs to this neighborhood, also any $\lambda 
{\bf V}$, with $\lambda \in \R$ and $0\leq \lambda\leq 1$, belongs to the 
neighborhood\footnote{It is possible to give a 
stronger definition requiring $\lambda \in \C$ and $|\lambda|\leq 1$. 
Anyway, throughout this paper 
we use the weaker definition.}. For instance, any complex open ball of 
centered in $z\in \C$ 
with positive 
radius $r>|z|$ is an open  starshaped neighborhood of the origin of
$\C$.
We can take each $E_{p}$ above  as  fixed  open 
starshaped  neighborhoods of ${\bf 0}$ where the 
exponential map centered on $p$ define a diffeomorphism.
Working in fixed local coordinates, it is trivially possible to choose such 
sets $E_p$ such that $E$ given in (\ref{E}) is also open.  Hence, the map
\beq
\phi : (p,{\bf X}) \mapsto (p, exp {\bf X}) \label{phi0}\:,
\eeq
defines a diffeomorphism in $E$ onto $\phi(E) \subset {\cal 
M}\times{\cal M}$ because it is injective and its differential does not
vanish in each point of the domain.\\
As usual, a {\em normal}
neighborhood of the point $p\in {\cal M}$, is an open neighborhood of 
$p$ with the form ${\cal N}_{p} = exp_p( S)$ whenever $S\subset E_p\subset
T_{p}({\cal M}) $ is an open  starshaped 
neighborhood of the origin of $T_p({\cal M})$
where the exponential map centered in $p$  defines an analytic diffeomorphism.
Then, the  components of the vectors ${\bf V}\in T_p({\cal M})$, with 
respects to a fixed base, contained in
$S$, define {\em  normal coordinates}  on ${\cal M}$ centered in  $p$
via the function ${\bf V}\mapsto exp_p {\bf V}$.
Notice that any $q\in {\cal N}_p$, due to (\ref{scaling}) and the 
starshapedness of $exp^{-1}_{p}{(\cal N)_{p}}$,
can be connected with $p$ by only one complex geodesic segment
 "starting from $p$" at $t=0$
and "terminating in $q$" at $t=1$, such that the associated real-parameter 
geodesic segment   is completely contained in ${\cal N}_p$. (Using the stronger
definition of starshaped neighborhood suggested in the previous footnote,
the whole complex segment geodesic would be contained in ${\cal N}_p$.)   
Finally, in normal coordinates        centered 
in $p$, due to (\ref{scaling}), the equation of a complex 
geodesic which starts form $p$ is a linear function of the parameter. 
This involves that the connection 
coefficients vanishes at $p$ if evaluated in these coordinates.\\
Similarly to the real case, 
we define a {\em totally normal neighborhood} of a point $p\in {\cal M}$
 as a  neighborhood \footnote{In this work,
a neighborhood
of a point is any set which includes an open set which contains
the point.} of $p$,
${\cal V}_p \subset {\cal M}$,
 such that,  if $q\in {\cal V}_p$, there is a normal neighborhood of $q$,
${\cal N}_q$, with ${\cal V}_p\subset {\cal N}_q$.
Therefore, if $q$ and $q'$
belong to the same totally normal neighborhood, there is only one
 complex geodesic segment which "connects" these two points (respectively for 
 $t=0$ and $t=1$) such that the associated real-parameter geodesic segment
is  completely
contained in normal neighborhoods centered in $q$ and $q'$ respectively,
 ${\cal N}_q$ and ${\cal N}_{q'}$ .\\
 Finally, 
 a {\em complex geodesically convex}
  neighborhood of a point $p\in {\cal M}$ should
be defined 
 as a totally normal neighborhood of $p$,
 ${\cal U}_p$, such that, for any couple $q,q'\in {\cal U}_p$,
there is only one complex geodesic segment which is completely
 contained in ${\cal U}_p$ and "connects" $q$ (for $t=0$) and  $q'$ 
 (for $t=1$). In fact, also in the simplest  case of a complex manifold 
 ${\cal M} \subset \C$
these neighborhoods do not exist barring trivial cases 
(e.g., ${\cal U}_p= {\cal M}=\C$). However, a weaker 
definition can successfully be given.
A {\em geodesically linearly convex} 
neighborhood of a point $p\in {\cal M}$ is defined
 as a totally normal neighborhood of $p$,
 ${\cal U}_p$, such that, for any couple $q,q'\in {\cal U}_p$,
there is only one {\em real-parameter} geodesic segment which is completely
 contained in ${\cal U}_p$ and connects $q$ (for $s=0$) and  $q'$
 (for $s=1$).\\
It is not so obvious, at this point,  that our
 complex pseudo-Riemannian structure admits 
totally normal and geodesically linearly convex neighborhoods. 
Actually this is the case.\\

\noindent {\bf Theorem 2.3.} {\em Let $({\cal M},{\bf g})$ be a complex 
pseudo-Riemannian manifold. For each point $p\in{\cal M}$ there is a local 
base of the topology $\{{\cal G}_{pj}\}_{j\in \R}$
 consisting of open totally normal, 
geodesically linearly convex neighborhoods of the point $p$. Moreover
 each $\bar {\cal G}_{pj}$ is also totally normal and 
geodesically linearly
convex for any $j\in \R$ and $\bar{{\cal G}}_{pj} \subset {\cal G}_{pj'}$
if $j<j'$.}\\

\noindent {\em Proof.} See the {\bf Appendix}. $\:\: \Box$\\

\noindent {\em Remark.} The definition of linearly geodesically convex 
neighborhoods could not seem very natural. Anyway, it can be given in a 
 more natural way for open sets (see also \cite{lb}). 
To this end, 
we leave to the reader the proof of the following relevant proposition.\\

\noindent {\bf Proposition 2.1.} {\em Given a complex pseudo-Riemannian 
manifold 
$({\cal M}, {\bf g})$, an {\em open} set ${\cal U}\subset
 {\cal M}$ is linearly 
geodesically convex, if and only if there is an open set $E({\cal U})
\subset T({\cal M})$, with
\beq
E({\cal U}) = \bigcup_{p\in {\cal U}} \{p\}\times E({\cal U})_p\:,
\eeq
$E({\cal U})_p\subset E_p$ being an open starshaped neighborhood of 
the origin of $T_p({\cal M})$, such that the map
\beq
\phi_{\cal U} : E({\cal U}) \to {\cal U}\times {\cal U}:
  (p,{\bf X}) \mapsto (p, exp_p {\bf X})
\eeq
is an analytic  diffeomorphism {\em onto} ${\cal U}\times{\cal U}$.}\\

\noindent The existence of totally normal and linearly 
geodesically convex neighborhoods allow us to define 
the {\em one half squared complex pseudo-distance} or {\em complex world 
function} similarly 
to the case of real metrics. Given an open  
linearly geodesically convex neighborhood, 
or, more simply, an open totally normal neighborhood,
${\cal U}$, the {\em complex world function} is given by
\beq
\sigma (p,q) := \frac{1}{2} {\bf 
g}(p)(exp_{p}^{-1}(q),exp_{p}^{-1}(q)) \:\:\:\: \mbox{for any }\:\: 
q,p\in {\cal U} \label{sigma}\:.
\eeq
Since all functions involved on 
the right-hand side of (\ref{sigma}) are analytic, it must hold
  $\sigma \in C^\omega({\cal U}\times {\cal 
U})$. Moreover, essentially from the conservation of 
${\bf g}({\bf V}, {\bf V})$ along any complex geodesic segment due to the 
geodesical transport,
${\bf V}$ being the tangent vector, we have the following 
properties which generalize
 well-known Riemannian and Lorentzian results \cite{fu}. 
\beq 
\sigma(p,q) &=& \sigma (q,p)\:,\label{sigma1} \\
 \sigma(p,q)  &=&\frac{1}{2}
\nabla_{(p)a}\sigma(p,q) \nabla_{(p)}^{a}\sigma(p,q)
\:,\label{sigma2} \\
\nabla_{(p)} \sigma(p,q) &=& \frac{d\:\:}{dt}|_{t=1} \gamma(t,q,p)
\:,\label{sigma3} 
\eeq
where the function on the right hand side of (\ref{sigma3}) is defined below.
In an open 
geodesically linearly convex neighborhood  or, more simply, in 
an open totally normal neighborhood, ${\cal U}$, we can define the function 
\beq
\gamma(t,p,q) := \gamma(t,p,exp_{p}^{-1}(q)) \:\:\:\: \mbox{for any }\:\: 
q,p\in {\cal U} \:\:\:\: \mbox{and }\:\: t\in 
\bar{B}_{1}(0)\label{gamma2}\:,
\eeq
which gives the complex geodesic segment "connecting" the point $p$ ($t=0$)
and the point $q$ ($t=1$) 
as a function of the extreme points. Once again, trivially,
$\gamma \in C^\omega( \bar{B}_{1}(0)\times {\cal U}\times {\cal U})$.\\
The property
(\ref{sigma2}) is a consequence of the property (\ref{sigma3}). 
The latter is not very simple to prove. A direct way is the following. 
 Consider a {\em normal coordinates system} centered in $q$. 
In these coordinates, if $\vec{x} = \vec{x}(t) (= t\vec{x}(1))$ 
is the equation of the real-parameter 
geodesic segment from $q\equiv \vec{x}(0) = \vec{0}$ to $p \equiv \vec{x}(1)$,
it holds trivially, since the integrand actually does not depend on $t$ due to
the parallel transport, $\sigma(p,q) = \frac{1}{2}
\int_0^1  g_{ab}(x(t)) \frac{\:dx^{a}}{dt}\frac{\:dx^{b}}{dt} dt$. 
Then we can vary the curve in the integrand within any family
of (real-parameter segment) 
geodesics  $\vec{x}_\alpha = \vec{x}_\alpha
(t)$, 
$t\in ]-\delta,\delta[$. Assume also that the dependence on $\alpha$ is 
smooth,  $\vec{x}_0(t) := \vec{x}(t)$ and 
$\vec{x}_\alpha(0) = \vec{0}$ for any $\alpha$.
This defines a functional $\sigma = \sigma[\vec{x}_\alpha]$. Using the
equation of the geodesic for $\alpha=0$, it is quite trivial to get by 
integration by parts that
$\frac{d \sigma[\vec{x}_\alpha]}{d\alpha}|_{\alpha=0} =
 g_{ab}(\vec{x}(1)) x^{a}(1)\frac{dx_\alpha^{b}(1)}{d\alpha}|_{\alpha=0}$. 
On the other hand, since each curve of the family is a geodesic and
$p\equiv \vec{x}(1)$, we have
$\frac{d \sigma[\vec{x}_\alpha]}{d\alpha}|_{\alpha=0} = 
\frac{d\sigma(\vec{x}_\alpha(1), \vec{0})}{d\alpha}|_{\alpha=0}
= {\partial}_{(p) b} 
\sigma(p,q) \frac{dx_\alpha^{b}(1)}{d\alpha}|_{\alpha=0}$.
Noticing that $\frac{d \vec{x}_\alpha(1)}{d\alpha}|_{\alpha=0}$ is arbitrary, 
one has (\ref{sigma3}).\\
Finally, let us consider the bi-scalar called 
 van Vleck-Morette determinant. In a real manifold
either Riemannian or Lorentzian, the definition (\ref{vvmdef}) can be 
rewritten, employing
 any coordinate system $\vec{z} = (z^1,\cdots,z^D)$ defined in an open
totally normal (or geodesically convex) neighborhood
${\cal T}$
 as
\beq
\Delta_{VVM}(x,y) := \frac{(-1)^D}{g(\vec{x})} 
\sqrt{\frac{g(\vec{x})}{g(\vec{y})}}
\det \left( \frac{\partial^2 \sigma(\vec{x},\vec{y})}{\partial x^{a} 
\partial y^{b}}
 \right) \label{VVMC}
 \:.
\eeq
This expression can be generalized to open totally normal neighborhoods 
in complex pseudo-Riemannian manifolds. Notice that the {\em bi-scalar}
 so obtained
is (complex) jointly-analytic in 
${\cal T}\times {\cal T}$, 
but, in principle,  can be a {\em multiple-valued} function due to the squared 
root. In any case,
the branch point of the squared root is harmless since 
 $g(\vec{z})\neq 0$. 
Computing   $\Delta_{VVM}(x,y)$
in normal coordinates centered in $x$  (these coordinates do exist
and cover ${\cal T}$ in our hypotheses), making use of (\ref{sigma3}), we get
that, {\em in these coordinates},
\beq 
\Delta_{VVM}(x,y) =
\sqrt{\frac{g(\vec{x})}{g(\vec{y})}} \:(\neq 0)\:. \label{VVMN}
\eeq
Therefore, the bi-scalar $\Delta_{VVM}(x,y)$ cannot vanish anywhere
and is positive either for a Riemannian or Lorentzian metric 
(all that not depending on the used coordinates!).\\

We are now able to state and prove the most important theorem for
our goal (we omit the index $\lambda$ in some notation 
 for the sake of simplicity).\\

\noindent {\bf Theorem 2.4.} 
{\em Let $({\cal M}, {\bf g})$ be
a $(D=d+1)$-dimensional Lorentzian manifold with class
$C^\omega$.
 Let  ${\cal O}\subset {\cal M}$ be any open set
endowed with ($C^\omega$) local synchronous coordinates (with respect 
to some spacelike hypersurface)
 $\vec{x} =
(x^0,\cdots,x^d)$. 
 Fix a positive real $\lambda$ and
 consider the set of complex pseudometrics $\{{\bf g}_{\lambda\theta}\}$ 
defined in} (\ref{otimes}) of
{\em {\bf Theorem 2.1}  in the analytically extended coordinates 
$z^0,\cdots,z^d$ ($z^a = x^a +i y^a$)
varying  in a open complex set ${\cal G}\subset \C^D$
with ${\cal O}\subset {\cal G}$} and $\theta \in {\C}$.

(a) {\em For any $p\in {\cal G}$, 
there is a local base of the topology of ${\cal G}$, 
$\{{\cal G}_{pj}\}_{j\in \R}$,
consisting of  open totally normal, geodesically linearly convex 
 neighborhoods of $p$
 {\em in common with}
 all of the complex pseudo-metrics ${\bf g}_{(\lambda\theta)}$ for
$\theta$ which belongs to an open complex neighborhood of $[0,\pi]$, 
${\cal K}_p$.
Moreover, each $\bar {\cal G}_{pj}$ is also totally normal and  
geodesically linearly convex, with respect to all of the complex pseudometrics 
when $\theta \in {\cal K}_p$,
and $\bar{{\cal G}}_{pj} \subset {\cal G}_{pj'}$
if $j<j'$.}

(b) {\em If $p\in {\cal O}$,   posing 
 (with obvious notations referred to 
the coordinates $\vec{z}$)
${\cal U}_{pj}:= Re\: {\cal G}_{pj}$, $\{{\cal U}_{pj}\}_{j\in \R}$,
is a local base of the topology of ${\cal O}$ about $p$, 
consisting of  open totally normal, 
 geodesically convex  neighborhoods of the point $p$ in common with 
whichever real (Riemannian or Lorentzian)  metric  produced, 
in the considered coordinates, 
by ${\bf g}_{(\lambda \theta)}(\vec{x})$ for particular choices of the, 
generally complex, value of $\theta\in {\cal K}_{p}$. 
In particular, this holds for the initial Lorentzian metric ${\bf g}$ 
($\theta=0$)
and for the final Riemannian Wick-rotated 
 metric $\bar{\bf g}_{\lambda}$ ($\theta = \pi$).
Moreover, each $\bar {\cal U}_{pj}$ is also totally normal and  
geodesically convex, with respect all of the real metrics considered above 
and $\bar{\cal U}_{pj} \subset {\cal U}_{pj'}$,
if $j<j'$.}

(c) {\em Arbitrarily fixed an element ${\cal G}_{pj}$,
the {\em complex} 
functions obtained from} (\ref{sigma}), (\ref{VVMC}) and (\ref{gamma2}) 
{\em  specialized to the generic complex 
pseudometric metric ${\bf g}_{(\lambda\theta)}$,}
\beq
(\theta, q, q')&\mapsto& \sigma_\theta(q,q')\:\:\:\:\mbox{{\em for}}\:\:\:\:
 (\theta,q,q')\in 
{\cal K}_p\times {\cal G}_{pj}\times {\cal G}_{pj} \label{sigmatheta}\\
(\theta, q, q')&\mapsto& \Delta^{1/2}_{VVM\theta}(q,q')
\:\:\:\:\mbox{{\em for}}\:\:\:\:
 (\theta,q,q')\in 
{\cal K}_p\times {\cal G}_{pj}\times {\cal G}_{pj} \label{deltatheta}\\
(\theta,t, q, q')&\mapsto& \gamma_\theta(t,q,q')
\:\:\:\: \mbox{{\em for}}\:\:\:\:
(\theta,t,q,q')\in 
{\cal K}_p\times B_2(0) \times 
{\cal G}_{pj}\times {\cal G}_{pj} \label{gammatheta}
\eeq
{\em are jointly-analytic functions. Moreover,  
$\Delta^{1/2}_{VVM\theta}(\vec{x},\vec{y})$ is
a {\em single-valued} function and can be defined such that 
it coincides with the usual real positive
van Vleck-Morette determinant for real metrics considered in} (b).\\

\noindent {\em Proof.} See the {\bf Appendix}. $\:\:\:\:\:\:\Box$

\section{The symmetry of Seeley-deWitt coefficients in smooth manifolds.}

\noindent {\bf 3.1.}  {\em The analytic case.} Let us consider the
off-diagonal 
Seeley-deWitt coefficients given in {\bf Definition 1.1}. 
It is possible to show that,
if the Lorentzian metric and the function $V$ are real analytic functions
 of the local coordinates, then the coefficients are symmetric functions
of the arguments $x$ and $y$. The way is direct, we can use the local
Wick rotation previously defined and, via {\bf Theorem 2.4}, we get the
symmetry of the Seeley-deWitt coefficients from the symmetry of heat-kernel
coefficients defined with respect the Wick-rotated Riemannian metric.\\

\noindent {\bf Theorem 3.1.} {\em Let $({\cal M},{\bf g})$ be a (real, 
Hausdorff, paracompact,
connected, orientable) $(D=d+1)$-Lorentzian $C^\omega$ manifold. Suppose 
the function $V$ which appears in} (\ref{caseL}) {\em is a
 (real) analytic function and consider
the Seeley-deWitt/Hadamard  coefficients given in {\bf Definition 1.1}.}\\
{\em Then, any point $p\in {\cal M}$ admits  a (totally normal)
geodesically convex neighborhood ${\cal N}_p$ 
such that, if $x,y\in {\cal N}_p$,
\beq
a_j(x,y) = a_j(y,x)\:, \label{symmetry} 
\eeq
for any $j\in \N$.}\\

\noindent {\em Proof.} Fix any point $p\in {\cal M}$ and consider
a synchronous coordinate system $x^{0,\ldots},x^d$ 
defined in a open neighborhood ${\cal O}$
of $p$. Fix $\lambda=1$ and 
 use {\bf Theorem 2.4} in ${\cal O}$ with respect to the coordinates 
$\vec{x}$. From now on, we shall use the notations
of {\bf Theorem 2.4}. Consider the local complex extension 
of the manifold defined on ${\cal G}$
and fix a common geodesically linearly convex set 
${\cal H}_p = {\cal G}_{pj_0}$ of the local base of the topology found 
in (a) of {\bf Theorem 2.4}. The Seeley-deWitt coefficients defined in 
${\cal N}_p := Re\: {\cal H}_p$ by {\bf Definition 1.1} can be analytically 
continued in the whole set ${\cal H}_p$. In particular we have that from
{\bf Definition 1.1} and (c) of {\bf Theorem 2.4},
fixing the index $j$,
and $x,y\in {\cal N }_p$, each function (with obvious notations)
\beq
\theta \mapsto 
a_j(\vec{x},\vec{y}|{\bf g}_{\theta})-a_j(\vec{y},\vec{x}|{\bf g}_{\theta})
\label{symm}\:,
\eeq
is analytic for $\theta \in {\cal K}_p$ where ${\cal K}_p$ is a  complex
open neighborhood of $[0,\pi]$. ${\cal K}_p$ can be assumed
to be open and connected dropping the connected components which do not
 contain $[0,\pi]$. 
In particular, we can consider the complex values
of $\theta$,
$\theta = \pi + i \mu$ where $\mu$ ranges in $[0,\epsilon[$. If $\epsilon$ 
is small enough, all these values of $\theta$ belong to ${\cal K}_p$.
Then, we notice that, using the notations defined in 
(\ref{analytic1}) and (\ref{analytic2}), we have
\beq
{\bf g}_{(\lambda=1 \:,\:\theta = \pi + i\mu)} = 
{\bf g}_{(\lambda = e^{-\mu/2}\:,\: \theta = \pi)} \:.
\eeq
The metric on the right hand side is {\em Riemannian}.

Since the analytical continuation of the metric preserves 
the form of the right-hand side of (\ref{s}),
 the analytical continuation  of the off-diagonal 
Seeley-deWitt coefficients of the initial
Lorentzian metric ${\bf g}$, for $\theta = \pi +i\mu$ 
produces the off-diagonal 
heat-kernel coefficients of the corresponding Riemannian metrics.
Therefore, as we know by \cite{m3}, the right hand side 
of (\ref{symm}) vanishes, whenever $\theta$ belongs to
 the set $\{\theta = \pi +i \mu\:|\: \mu \in [0,\epsilon [\}
\subset {\cal K}_p$ 
for some $\epsilon >0$. The uniqueness of the analytic continuation 
in open connected sets
entails that the right hand side of (\ref{symm}) vanishes everywhere
in ${\cal K}_p$, and in particular for $\theta = 0$. This means that
the off-diagonal
Seeley-deWitt coefficients defined with respect to the initial metric
are symmetric functions of $x$ and $y$ in ${\cal N}_p$. $\:\:\:\:\:\:\Box$\\

\noindent The result just proved implies the following 
 more general result in a direct way,
as remarked in {\bf 2.2} of \cite{m3}.\\

\noindent{\bf Theorem 3.2.}  {\em Let $(M,{\bf g})$ be a (real, 
Hausdorff, paracompact,
connected, orientable) $(D=d+1)$-Lorentzian $C^\infty$ manifold.
 Consider
the Seeley-deWitt Hadamard  coefficients given in {\bf Definitions 1.1} when
both the metric ${\bf g}$ and the function $V$ which appears in} (\ref{caseL}) 
{\em  are smooth fields.}\\
{\em Then, any point $p\in {\cal M}$ admits  a (totally normal)
geodesically convex neighborhood ${\cal N}_p$ such that,
 if $x,y\in {\cal N}_p$,
\beq
a_j(x,y) = a_j(y,x)\:, \label{symmetry2} 
\eeq
for any $j\in \N$.}\\

\noindent {\em Proof.} The proof is exactly the same performed in the 
Riemannian case, {\bf Theorem 2.2} in \cite{m3}. $\:\:\:\:\:\:\:\: \Box$\\

\noindent {\em Remark.} This result can be achieved also if the manifold admits
a smooth boundary as pointed out in \cite{m3}.\\

Finally, we have a trivial corollary based on the fact that the Seeley-deWitt 
coefficients are also  the coefficients which appear in the Hadamard local 
solution \cite{m3}.\\

\noindent  {\bf Corollary of Theorem 3.2.} 
{\em Let $(M,{\bf g})$ be a (real, 
Hausdorff, paracompact,
connected, orientable) $(D=d+1)$-Lorentzian $C^\infty$ manifold.
Let the metric ${\bf g}$ 
and the function $V$ in} (\ref{caseL}) {\em be smooth fields.}\\
{\em Then, for any point $p\in {\cal M}$
 there is a (totally normal)
 geodesically convex neighborhood ${\cal N}_p$  of $p$,
such that, for any pair $(x,y)\in {\cal N}_p$,
 the coefficients $u_j,v_j$ of the Hadamard parametrix,
 up to the order indicated (see} (22) {\em and} (23) 
{\em in} \cite{m3}{\em ),
\begin{eqnarray}
H_N(x,y) &=& \sum_{j=0}^{D/2-2}
\left(\frac{2}{\sigma(x,y)}\right)^{D/2-j-1} u_j(x,y) +
\sum_{j=0}^{N} \sigma^j(x,y) v_j(x,y) \ln (\sigma(x,y)/2) \label{V}
\end{eqnarray}
($N\in \N$ fixed arbitrarily)
for $D$ even (the former summation appears for $D\geq 4$ only), and
\begin{eqnarray}
H(x,y) &=& \sum_{j=0}^{(D-5)/2}
\left(\frac{2}{\sigma(x,y)}\right)^{D/2-j-1} u_j(x,y) + v_0(x,y)
\sqrt{\frac{2\pi}{\sigma(x,y)}} \nonumber\\
& &+ v_1(x,y) \sqrt{2\pi \sigma(x,y)}
\end{eqnarray}
for $D$ odd (the summation  appears for $D\geq 5$ only), satisfy
\begin{eqnarray}
u_j(x,y) &=&  u_j(y,x)\:, \\
v_j(x,y) &=&  v_j(y,x)\:.
\end{eqnarray}
}

\noindent {\bf 3.2.} {\em Final remarks.} The results proved in this work
show that the Seeley-deWitt Hadamard  off-diagonal  coefficients 
are symmetric as requested within the point-splitting 
renormalization procedure of the stress-energy tensor. Such a result 
has been proved  for the case where the 
manifold, the metric and the potential $V$ are smooth. 
The result  holds true in both  Lorentzian and Riemannian  manifolds.  
Anyway, the intriguing general 
fact we have pointed out is the existence of  quite a natural local 
Wick rotation of the metric which preserves the local geodesical structures
of the manifold making use of non-hermitian complex manifolds. This
procedure makes sense regardless the presence of time-like Killing fields
whenever the employed coordinates are somehow "physical".
It is not so obvious  what physics is involved in this procedure.\\

\noindent {\em Acknowledgement.} I am grateful to A.Cassa and S.Baldo,
S.Delladio, S.Hollands, A.Tognoli and P.Vigna Suria for very useful 
remarks, suggestions and discussions. \\
A part of this work has been written during my visit at the Department of
Mathematics of the University of York. I would like to thank 
C.J. Fewster, A. Higuchi, and Bernard S. Kay in particular, 
for very stimulating discussions and for the cordial hospitality 
all they provided during my stay there.\\ 
This work has been financially supported by both a postdoctoral fellowship 
of the Department of Mathematics of the University of Trento
and a grant by the MURST within the National Project "Young Researchers".

\section*{Appendix: Proof of some Theorems and Lemmata.}

\noindent{\em Sketch of Proof of {\bf Theorem 2.2}.}\\
 Fix a point $p\in {\cal S}$. 
Since ${\cal S}$ is embedded, it is possible to find a local 
coordinate system centered in $p$, $\vec{x} = (x^{0},x^{1}\cdots, x^d)$, 
defining a local chart  $({\cal U}_{p}, \vec{x})$
\ about $p$ such that the set ${\cal 
U}_p\cap {\cal S}$ is given by the equation $x^{0} = 0$. Then 
$(x^{1},\cdots,x^d)$ define local {\em space-like} 
coordinates on ${\cal S}$ in a neighborhood of $p$. Now consider the 
local map $(t,x^1,\cdots, x^d) \mapsto exp_{(0,x^{1},\cdots,x^d)}(t {\bf 
N}(x^{1},\cdots,x^d))$ which is defined in an open neighborhood of 
$(t=0,x^{1}= 0,\cdots,x^d =0)$. ${\bf N}(x^{1},\cdots,x^d)$ is
the unique time-oriented vector normal to ${\cal S}$ 
in $(x^{0}=0,x^{1},\cdots, x^d)$ with 
${\bf g}({\bf N},{\bf N})=-1$. It is a trivial task to compute the 
Jacobian determinant $J_{p}$ of the map $(t,x^1,\cdots, x^d) 
\mapsto \vec{x}^{-1}
\circ exp_{(0,x^{1},\cdots,x^d)}(t {\bf N}(x^{1},\cdots,x^d))$
at $(t=0,x^{1}= 0,\cdots,x^d =0)$ obtaining $J_{p} =
dx^{0}|_{p}({\bf N}(p)) \neq 0$.  Hence, the map  $(t,x^1,\cdots, x^d) 
\mapsto exp_{(0,x^{1},\cdots,x^d)}(t {\bf N}(x^{1},\cdots,x^d))$
is a coordinate system in an open neighborhood of $p$, 
${\cal U}'_{p}\subset {\cal U}_p$ and
 $(y^{0},y^{1},\cdots,y^d) := 
 (t,x^{1},\cdots, x^d)$ are local coordinates about $p\in {\cal S}$. 
 Using the equation of geodesics, it is a trivial task to get  that 
 (\ref{Fermi1}) and (\ref{Fermi2}) 
 are fulfilled and thus (making smaller ${\cal U}_{p}$, if necessary,
 in order to have a $\vec{y}$ domain of the form 
 $\:]-\delta,\delta[\:\times
 \Omega\:$) we have built up a (time-oriented) 
 locally synchronous coordinates with respect to $\cal S$. So local synchronous 
 coordinates do exist.  On the other hand it is also simply proved
 that the temporal coordinate of a point $q$ in any (time-oriented)
local synchronized coordinate system defined in {\bf Definition 2.1}
represents the (positive) length $t_{q}$ of the {\em unique} 
geodesic segment which starts 
from ${\cal S}$ with a unitary  initial tangent vector time-oriented 
normal to ${\cal S}$ at, say, $q'\in {\cal S}$ and reaches $q$. 
The spatial synchronous  coordinates are nothing but the 
coordinates of $q'$ on ${\cal S}$. Then,
{\bf Proposition 26} in Chap.7 of \cite{oneill} entails that 
there is an open neighborhood ${\cal O}$ 
of ${\cal S}$ where any pair of geodesics starting from different 
points of ${\cal S}$ with  initial 
tangent vector normal to ${\cal S}$ do not intersect each other 
anywhere (also if the starting points belong to {\em different}
local synchronous coordinate system domains). 
By consequence, in
 ${\cal O}$,  the temporal coordinate $q\mapsto t_{q}$ of any point $q$ 
does {\em not} depend on the chosen local synchronous coordinate 
system. The coordinate transformation 
law between local synchronous coordinate system reads, in any common domain,
\beq
{y'}_{q}^0 &=& y_{q}^{0} = t_{q} \label{simple1} \:,\\
{y'}_{q}^j & = & {y'}_{q}^j(y_{q}^{1}, \cdots, y_{q}^d) \:\:,\:\: j= 1,\cdots,d 
\label{simple2}\:.
\eeq
This trivially assures that the transformation law from different local 
synchronous coordinates preserves the form of the Wick rotated metric 
in common domains for any globally fixed value of $\lambda$.
This defines a Riemannian metric on ${\cal N}$ which can be taken as the 
union of all af the intersections of ${\cal O}$ with each synchronous 
chart domain. 

$\:\:\:\:\: \:\:\Box$\\

\noindent{\em Proof of {\bf Lemma 2.1}.}\\
The differential equation system in {\bf Lemma 2.1} is equivalent
to the integral equation
\begin{eqnarray}
Y(z,\bar{y}_0,\alpha) = 
\bar{y}_0 + \int_{z_0}^{z} f(u,Y(u,\bar{y}_0,\alpha),\alpha) du \label{du}
\end{eqnarray}
where the path of integration is the segment from $z_0$ to $z$
($Y$ is $C^1$ and thus analytic in $z$ and the integration does not 
depend on the chosen path between the same extreme points). We can write the
equation above as
\begin{eqnarray}
Y = A_{\bar{y}_0\alpha}(Y)
\end{eqnarray}
where $A_{\bar{y}_0\alpha}$ is defined by the right-hand side of
(\ref{du}) and it should be thought as a  function  which maps
 the Banach space 
$B:= C^0(\bar{B}_{r_1}(z_0);\C^n)$ (with the norm  $||\:\:\:||_{\infty}$)
into $B$ itself. Actually  $f(z,Y(z),\alpha)$ may  not be
defined, in general,  when $Y\in B$  because some
$Y(z)$ may be out of the domain of $f$.
However, once one has fixed $r'_2>0$ such that $r'_2<r_2$,
taking a value $r'_1>0$ which satisfies
(\ref{delta})
one sees that $A_{\bar{y}_0\alpha}$ is well-defined on the
the closed subset of $B$,
\begin{eqnarray}
B_0 := \{ Y \in C^0(\bar{B}_{r_1}(z_0);\C^n)
 \:\:|\:\: Y(z) \in \bar{B}_{r_2}(y_0)
\:\: \mbox{\em{for any}}\:\:z \}
\end{eqnarray}
which is  invariant
under $A_{\bar{y}_0\alpha}$ provided
$\bar{y}_0 \in \bar{B}_{r'_2}(y_0)$. In this domain,
 $A_{\bar{y}_0\alpha}$ is a contraction map, with  contraction 
constant $\rho$, such that  $0< \rho<1 $ which does  {\em not} depend 
on $\bar{y}_0\in \bar{B}_{r'_2}(y_0)$ and $\alpha\in \bar{B}_{r_3}(\alpha_0)$,
on that set.
Banach  theorem of the
fixed point proves the existence and the uniqueness of the solution which is
nothing but the fixed point of $A_{\bar{y}_0,\alpha}$ and belongs to $B_0$.
In particular, the solution can be found as the limit 
(in the norm $||\:\:||_\infty$ with respect to the variable $z$, the 
remaining variables being fixed.)
\beq
Y = \lim_{k\to\infty} Y_{k} \label{seq}\:,
\eeq
where 
$Y_k := A^k_{\bar{y}_0\alpha}(Y_0)$ and
$Y_0$ is the constant function $z \mapsto Y_{0}(z) = y_0$ everywhere. Using
the contraction property one finds that, for $k>m$,
\beq
Sup ||Y_k(z,y,\alpha) -Y_m(z,y,\alpha) ||
\leq \frac{\rho^{k-m}}{1-\rho}\: Sup ||Y_1(z,y,\alpha) -Y_0(z,y,\alpha) ||\nonumber
\:,
\eeq
where the $Sup$ is evaluated for $(z,y,\alpha)\in
\bar{B}_{r'_{1}}(z_{0})\times \bar{B}_{r'_{2}}(y_{0})
 \times  \bar{B}_{r_{3}}(\alpha_{0})$.
This entails that the convergence of the sequence (\ref{seq}) is uniform 
in all variables jointly. Since  each function of the series is analytic by 
construction in any set $\bar{B}_{r'_1} \times \bar{B}_{r'_{2}}(y_{0})
 \times  \bar{B}_{r'_{3}}(\alpha_{0})$, $0<r'_3<r_3$, 
the limit function must be analytic therein. $\:\:\:\:\:\:\Box$\\

\noindent {\em Proof of {\bf Theorem 2.3}.} \\
We follow and generalize the
similar proof given in \cite{kn}. 
Let $n$ the dimension of ${\cal M}$. Take a coordinate
system centered in $p\in {\cal M}$, $\vec{z} = 
(z^{1},\cdots,z^{n})$, $p\equiv (0,\cdots,0)$.
 We want to show that in these coordinates it is possible to find a class
of open totally normal neighborhoods of $p$ of the form $B_{\rho}(\vec{0})
:= \{ \vec{z}\in \C^n\:\:|\:\: \sum_{i=1}^n |z^{i}|^{2} <\rho^{2}\}$, 
$0<\rho<\bar{\rho}$, which are also linear
geodesically convex and the class of the sets  $\bar{B}_{\rho}(\vec{0})$ 
enjoys the same properties. The 
remaining part of the thesis is trivially proven by defining, 
for a fixed $\rho$, $0<\rho<\bar{\rho}$,
 ${\cal G}_{pj}:= B_{\rho(1+\tanh j)/2}(\vec{0})$ with $j\in \R$.
The existence of the class above follows from a pair of propositions
indicated by (p1) and (p2) in the following.

(p1) {\em Let $S_{\rho}(\vec{0}) :=\{ \vec{z}\in \C^n\:\:|\:\: 
\sum_{i=1}^n |z^{i}|^{2} = \rho\} $, $\rho>0$, then there exists $c>0$  
such that if $\rho \in ]0, c[$, then any {\em real-parameter geodesic} 
which is tangent to 
$S_{\rho}(\vec{0})$ at a point, say $\vec{y}$, lies outside 
$S_{\rho}(\vec{0})$ in a neighborhood of $\vec{y}$}.\\

\noindent {\em Proof of} (p1). Let $\vec{z} =\vec{z}(s)$, $s$ defined in a
neighborhood of $s_0$, be 
a real-parameter geodesic which is tangent to $S_{\rho}(\vec{0})$ at
 $\vec{y} = \vec{z}(s_0)$ ($\rho$ will be restricted later). Formally 
 speaking, this means that, setting
 \beq
 F(s) := ||\vec{z}(s)||^{2} \label{F}\:,
 \eeq
 it holds
 $F(s_0) = \rho^{2}$ and 
\beq
\frac{dF}{ds}|_{s=s_0} = \sum_{i}
 z^{*i}(s_0)\frac{dz^{i}}{ds}|_{s=s_0} +  z^{i}(s_0) 
 \frac{dz^{*i}}{ds}|_{s=s_0} = 0\:. \label{F'}
\eeq
Let us consider the second derivative of $F$ at $s=s_0$.
A trivial computation based on the derivation of central term in 
(\ref{F'}) and the equation of geodesics shows that
\beq
\frac{d^{2}F}{d s^2}|_{s=s_0}= V^{\dagger} A(\vec{z}, \vec{z}^{*}) V 
\label{F''}
\eeq
 where $V$ is a vector with components $V^j = d z^j/ds|_{s=s_0}$ for
 $j= 1,\cdots n$ and $^{*}$ is the complex conjugation and $^\dagger$
 the hermitian conjugation.
 $A(\vec{z}, \vec{z}^{*})$ is a $2n\times 2n$ Hermitian matrix 
with
\beq
A(\vec{z}, \vec{z}^{*})_{ij} = \delta_{ij}
\eeq
for $i,j = 1,\cdots,n$ and $i,j = n+1,\cdots 2n$, and
\beq
A(\vec{z}, \vec{z}^{*})_{ij} =  - \sum_{a=1}^{n}z^a \Gamma^{a}_{j\:
k-n}(\vec{z})
\eeq
for $i=1,\cdots, n$ and $j= n+1,\cdots,2n$, and finally
\beq
A(\vec{z}, \vec{z}^{*})_{ij} =  - \sum_{a=1}^{n}z^{*a} \Gamma^{*a}_{j-n \:k}
(\vec{z})
\eeq
for $i= n+1,\cdots,2n$. and $j=1,\cdots, n$.
The matrix $A$ becomes the identity matrix for $\vec{z} =\vec{z}^{*} = 
\vec{0}$  and thus is positive definite. There is a neighborhood of 
$\vec{0}$ which can be chosen in the form of $B_{c}(\vec{0})$, where, 
by continuity, $A(\vec{z}, \vec{z}^{*})$ is defined positive. In this 
neighborhood $d^2F/ds^2|_{s=s_0}>0$ and hence $F(s) > \rho^2$ when $s\neq s_0$ 
belongs to a
 real neighborhood of $s_0$, provided $\rho \in ]0,c[$ \footnote{In general, 
this is not true in a {\em complex}
neighborhood of $0$ as one could trivially check in ${\cal M}=\C$.}. This conclude the proof of
(p1). $\:\:\:\:\:\:\Box$.\\

(p2) {\em Choose a real $c>0$ as in} (p1). 
{\em Then  there exists a real $a$ with 
$0<a<c$ such that:} 
(1) {\em Any two points of $B_{a}(\vec{0})$ ($\bar{B}_a(\vec{0})$)
 can be joined by a complex
geodesic segment which lies in $B_{c}(\vec{0})$;}
(2) {\em Each point of $B_{a}(\vec{0})$ ($\bar{B}_a(\vec{0})$)
has a normal coordinate neighborhood
containing $B_{a}(\vec{0})$ ($\bar{B}_{a}(\vec{0})$)
 and thus is a {\em totally normal} neighborhood.} \\

\noindent {\em Proof of} (p2). Consider ${\cal M}$ as a submanifold of 
$T({\cal M})$ in a natural way and work in coordinates $\vec{z}$ defined 
above, notice that  $p\equiv \vec{0}$. Set
\beq
\phi : {\bf X} \mapsto (q, exp{\bf X})\:\:\:\:\mbox{for}\:\:
 {\bf X}\in T_q({\cal M}) \label{phi}\:.
\eeq
In general, $\phi$ is defined only in a neighborhood of ${\cal M}$ in
 $T({\cal M})$. Since the differential of $\phi$ at $(\vec{0}\equiv p, 
{\bf 0})$
  is nonsingular, there exist a neighborhood $V$ of $(\vec{0},{\bf 0})$ 
in $T({\cal M})$
and a positive number $b<c$ such that $\phi$ defines a diffeomorphism
in $V$ onto $B_b(\vec{0})\times B_b(\vec{0})$. Taking $V$ and $b$
small, we can assume that $exp(t{\bf X})\subset B_c(\vec{0})$ for all
${\bf X}\in V$ and  $t\in \C$, $|t|\leq 1 $. The item (1) holds true
for any  $a>0$ with $a\leq b$, since
the complex geodesic segment from $q\in B_b(\vec{0})$ to $q'\in B_b(\vec{0})$
is the map $t\mapsto exp(t {\bf X})$ where $|t|\leq 0$ and 
${\bf X} := \phi^{-1}(q,q') \in V$.\\ 
Let us consider the item (2). Fixed the positive real $b$ and $V$ as those 
in the proof of item (1), choosing $b'>0$ and $\delta>0$ small enough,
 we can fix an open subset of $V$,  which is a neighborhood of $\vec{0}$ in
$T({\cal M})$, with the form $B_{b'}(\vec{0}) \times B_\delta$, where 
$0<b'< b $ and $B_\delta$ is an open ball of radius $\delta>0$   and
center in ${\bf 0}\in \C^n$. All the tangent spaces 
$T_q({\cal M})$, $q\in B_{b'}(\vec{0})$, have been 
identified with $\C^n$ by means of the bases induced by the considered
 coordinates. Then, choose an open neighborhood $B_{b''}(\vec{0})
\times B_{b''}(\vec{0}) \subset \phi(B_{b'}(\vec{0}) \times B_\delta)$.
Finally notice that, if $q\in B_{b''}(\vec{0})$, since $\phi$ 
is a diffeomorphism in $B_{b'}(\vec{0}) \times B_\delta$, we have
\beq
 \{q\}\times B_{b''}(\vec{0}) \subset \phi(\{q\}\times B_\delta) \nonumber\:,
\eeq
and in particular, from the definition of $\phi$,
\beq
B_{b''}(\vec{0}) \subset exp_q(B_\delta)\:.
\eeq
This means that $B_{a}(\vec{0})$ is a totally normal neighborhood if
$a\leq b''$ (and (1)  also holds true due to $b''<b$). \\
The proof for the closure of the considered neighborhoods  is trivial
and is obtained by taking $a$ smaller and noticing that 
$\bar{B}_{a'}(\vec{0}) \subset B_{a}(\vec {0})$ if $a'<a$.
$\:\:\:\:\:\:\Box$\\

\noindent
To complete the proof of the theorem, let $0<\rho<a (<c)$ and let $q,q'$ be
any pair of points in $B_\rho(p)$ ($\bar{B}_\rho(p)$). 
Let $\vec{z} = \vec{z}(s)$,
$s\in [0,1]$ the real-parameter segment geodesic from $q$ to $q'$ in
$B_{c}(p)$ (see (p1)). We shall show that this real-parameter
segment geodesic lies completely in $B_\rho(p)$ ($\bar{B}_\rho(p)$). 
Consider the function 
$s\mapsto F(s)$ defined in (\ref{F}) along this geodesic  segment. Assume that 
$F(s) \geq \rho^2$ 
($F(s) > \rho^2$ ) for some $s$, that is, $\vec{z}(s)$ lies outside 
$B_\rho(p)$ ($\bar{B}_\rho(p)$ )
for some $s$. Let $s_0$, $s_0\in ]0,1[$, be the value for which $F$
attains the maximum, say, $\rho_0^2\geq \rho^{2}$
 ($\rho_0^2> \rho$). Then
\beq
0= \frac{dF}{ds}|_{s=s_0} \:.
\eeq
This means that the real-parameter geodesic segment is tangent to the sphere
$S_{\rho_0}(p)$ at the point $\vec{x}(s_0)$. By the choice of $\rho$ the 
considered real-parameter geodesic segment lies inside the sphere
 $S_{\rho_0}(p)$, contradicting (p1). $\:\:\:\:\:\:\:\Box$\\

\noindent {\em Proof of {\bf Theorem 2.4}.} \\
First of all, we notice that the item (c) is a direct consequence of 
items (a) and (b), {\bf Lemma 2.1} and the discussion which follows that lemma.
Barring the  item (a), the only not completely trivial fact is the statement concerning 
the possibility of defining $\Delta_{VVM\theta}^{1/2}$  as a 
 single-valued function which coincides with the usual one
when evaluated on any ${\cal U}_{pj}$ for real (Riemannian or 
Lorentzian metrics). 
We shall prove this result in the end of the proof of this theorem.\\
Let us prove the  validity of items (a) and (b). The latter is a
straightforward consequence of the former taking into account that the 
the initial, the Wick-rotated and any other real metric obtained for the
corresponding values of $\theta$
when restricting to real coordinates produce real exponential maps end geodesics. 
(Therefore, with respect to the considered coordinates,
the exponential map transforms vectors with real components
onto points with real coordinates. And the real-parameter geodesic segments
connecting pairs of points in any  $Re \:{\cal G}_{pj}$ 
($\overline{Re \:{\cal G}_{pj}}$) 
are {\em real} geodesic segments completely contained in 
$Re \:{\cal G}_{pj}$, ($\overline{Re \:{\cal G}_{pj}}$).) 
 Then we have to prove item (a) only.
To this end, we use the same proof of {\bf Theorem 2.3}
with the necessary modifications.\\

\noindent {\em Proof of} (a).
Fix any $\lambda >0$. (From now on, for the sake of simplicity,
 we omit the index $\lambda$ where not strictly necessary.)
Take the complex coordinate system considered in the 
hypotheses. We are free to move the origin of the coordinate
in $p\in {\cal G}$ by a complex translation. Let $\vec{u}=(u^1,\cdots,u^D)$
the new coordinate system. Therefore $p\equiv (0,\cdots,0)$. 
 We want to show that, in these coordinates, it is possible to find a class of
open totally normal neighborhoods of $p$, corresponding to the metric 
${\bf g}_{(\lambda \theta)}$, of the form 
\beq
B_{\rho}(p)
:= \{ \vec{u}\in \C^D\:\:|\:\: \sum_{i=1}^D |u^{i}|^{2} <\rho^{2}\}\:,
\label{palla}
\eeq 
$0<\rho<\bar{\rho}$, which are also linear
geodesically convex and the class of the sets $\bar{B}_{\rho}(\vec{0})$
  enjoys the same properties.
Moreover all these properties of a {\em fixed} neighborhood are preserved 
{\em varying} $\theta$ in a complex open neighborhood of $[0,\pi]$, ${\cal K}_p$.
 The remaining part of the 
thesis is trivially proven by defining, for a fixed $\rho$ with 
$0<\rho<\bar{\rho}$,
\beq 
 {\cal G}_{pj}:= B_{(1+\tanh j) \rho/2}(\vec{p}) \label{palla'}\:,
\eeq
 with $j\in \R$. 
The thesis follows from a pair of  propositions indicated by (p1) and (p2)
in the following.

(p1) {\em Let $S_{\rho}(\vec{0}) :=\{ \vec{u}\in \C^D\:\:|\:\: 
\sum_{i=1}^D |u^{i}|^{2} = \rho\} $, $\rho>0$, then there exist $c>0$  
and a complex open neighborhood of $[0,\pi]$, ${\cal K}_p'$,
such that if $\rho \in ]0, c[$, then any {\em real-parameter geodesic} 
defined with respect to any fixed metric ${\bf g}_{(\lambda \theta)}$ with
$\theta\in {\cal K}_p'$  which is tangent to 
$S_{\rho}(\vec{0})$ at a point, say $\vec{y}_{\theta}$, lies outside 
$S_{\rho}(\vec{0})$ in a neighborhood of $\vec{y}_\theta$}.\\

\noindent {\em Proof of} (p1). Fix  $\theta \in [0,\pi]$ arbitrarily, 
let $\vec{u} =\vec{u}_\theta(s)$, 
$s$ being defined in a neighborhood of $s_{\theta0}$ 
and with respect to the metric 
${\bf g}_{(\lambda \theta)}$, be 
a real-parameter geodesic which is tangent to $S_{\rho_\theta}(\vec{0})$ at
 $\vec{y}_\theta 
= \vec{u}_\theta(0)$ ($\rho_\theta$ will be restricted later).
Formally 
 speaking, this means that, setting
 \beq
 F_\theta(s) := ||\vec{u}_\theta(s)||^{2} \label{Ftheta}\:,
 \eeq
 it holds
 $F_\theta(0) = \rho_\theta^{2}$ and 
\beq
\frac{dF_\theta}{ds}|_{s=s_{\theta0}} = \sum_{i}
 u_\theta^{*i}(s_{\theta0}) \frac{du_\theta^{i}}{ds}|_{s=s_{\theta0}} +  
u_\theta^{i}(s_{\theta0}) 
 \frac{du_\theta^{*i}}{ds}|_{s=s_{\theta0}} = 0\:. \label{Ftheta'}
\eeq
Let us consider the second derivative of $F_\theta$ at $s=s_{\theta0}$.
A trivial computation based on the derivation of central term in 
(\ref{Ftheta'}) and the equation of geodesics shows that
\beq
\frac{d^{2}F_\theta}{d s^2}|_{s=s_{\theta0}}= V_\theta^{\dagger} 
A_\theta(\vec{u}_\theta, \vec{u}_\theta^{*}) V_\theta 
\label{Ftheta''}
\eeq
 where $V_\theta$ is a vector with components $V_\theta^j = du_\theta^j/ds
|_{s=s_{\theta0}}$ for
 $j= 1,\cdots D$ and  $^{*}$ is the complex conjugation and $^\dagger$
 the hermitian conjugation.
 $A(\vec{u}_\theta, \vec{u}_\theta^{*})$ is a $2D\times 2D$ Hermitian
matrix 
with
\beq
A_\theta(\vec{u}, \vec{u}^{*})_{ij} = \delta_{ij}
\eeq
for $i,j = 1,\cdots,D$ and $i,j = D+1,\cdots 2D$, and
\beq
A_\theta(\vec{u}, \vec{u}^{*})_{ij} =
  - \sum_{a=1}^{D}u^a \Gamma^{a}_{(\theta) j \:k-D}(\vec{u})
\eeq
for $i=1,\cdots, D$ and $j= D+1,\cdots,2D$, and finally
\beq
A_\theta(\vec{u}, \vec{u}^{*})_{ij} =  
- \sum_{a=1}^{D}u^{*a} \Gamma^{*a}_{(\theta)j-D\: k}
(\vec{u})
\eeq
for $i= D+1,\cdots,2D$. and $j=1,\cdots, D$.
The matrix $A_\theta$ becomes the identity matrix for $\vec{u}=\vec{u}^{*}
= \vec{0}$  and thus is positive definite. 
Now we let the coefficient $\theta$ in $A_\theta$
vary in a neighborhood of the initial value and rename the variable $\theta$
by $\eta$.
Due to the joint continuity of the connection coefficients, 
there is an open neighborhood of 
$(\theta, \vec{0})$ 
which can be chosen in the form of $B_{\delta_\theta}(\theta)\times 
B_{c_\theta}(\vec{0})$, where
 $A_\eta(\vec{u}, \vec{u}^{*})$ is positive definited.
 In this 
neighborhood $d^2F_\eta/ds^2|_{s=s_{\theta0}}>0$ and hence $F_\eta(s) > 
\rho_\eta^2$ 
when $s\neq s_{\theta0}$ belongs to a real neighborhood of $s_{\theta0}$.
This procedure can be performed for any point $\theta \in [0,\pi]$
obtaining a covering of this  set made by complex open balls
$B_{\delta_\theta}(\theta)$. Since $[0,\pi]$ is compact also as a complex set,
we can extract a finite covering made of balls 
centered on the points $\theta_i$, $i= 1,\cdots,N$, whose union is a complex
open neighborhood of $[0,\pi]$, ${\cal K}_p'$. Let $c = Min
\{ c_{\theta_i}\:|\: 
i=1,\cdots,N \}$. 
If $\vec{z} \in B_c(\vec{0})$ and $\theta \in {\cal K}_p'$,
we have    
 $d^2F_\theta/ds^2|_{s=s_{\theta0}}>0$ and hence $F_\theta(s) > \rho_\theta^2$ 
when $s\neq s_{\theta0}$ belongs to a real neighborhood of $s_{\theta0}$, provided 
$\rho_\theta \in ]0,c[$.
 This conclude the proof of
(p1). $\:\:\:\:\:\:\Box$\\

(p2) {\em  Choose a real $c>0$ as in} (p1). {\em Then there exist a real 
$a$ with $0<a<c$ and a complex open neighborhood of $[0,\pi]$, ${\cal K}_p''$
 such that:} (1) {\em Fixing any $\theta\in {\cal K}_p''$,
  any two points of $B_a(\vec{0})$ ($\bar{B}_a(\vec{0})$)
can be joined by a complex geodesic segment of 
the metric ${\bf g}_{(\lambda \theta)}$
 which lies in  $B_c(\vec{0})$;} (2) {\em Fixing any $\theta \in {\cal K}_p''$,
each point of $B_a(\vec{0})$ ($\bar{B}_a(\vec{0})$) has a normal 
coordinate neighborhood,
with respect to the metric ${\bf g}_{(\lambda \theta)}$,
containing $B_a(\vec{0})$ ($\bar{B}_a(\vec{0})$), 
and thus   $B_a(\vec{0})$ ($\bar{B}_a(\vec{0})$) is
 a} totally normal neighborhood {\em with
respect to any metric}  ${\bf g}_{(\lambda \theta)}$.\\
 
\noindent {\em Proof of} (p2). Consider ${\cal G}$ as a submanifold of 
$T({\cal G})$ in a natural way and work in coordinates $\vec{u}$ defined 
above, notice that  $p\equiv \vec{0}$. Set
\beq
\Phi : (\theta, {\bf X}) \mapsto (\theta, q, exp_{(\theta)}{\bf X})
\:\:\:\:\mbox{for}\:\:
 {\bf X}\in T_q({\cal M}), \theta \in {\C}  \label{Phi}\:.
\eeq
In general, $\Phi$ is defined only in a neighborhood of ${\cal G}$ in
 $T({\cal G})$. Since the differential of 
$\Phi$ at $(\theta, \vec{0}, {\bf 0})$
for $\theta\in [0,\pi]$  is nonsingular, there exist an open neighborhood 
$V_\theta$ of $(\theta,\vec{0},{\bf 0})$ in $\C\times T({\cal G})$, a complex 
open neighborhood $B_{r_{\theta}}(\theta)$ of $\theta$
and a positive number $b_\theta<c$ such that $\Phi$ defines a diffeomorphism
in $V_\theta$ onto $B_{r_{\theta}}(\theta)\times B_{b_\theta}(\vec{0})\times
B_{b_\theta}(\vec{0})$. 
Taking $V_\theta$, $r_\theta$
and $b_\theta$ small, we can assume that $exp_{(\eta)}(t{\bf X})\subset 
B_c(\vec{0})$ for all
${\bf X}\in V_\theta$, $t\in \C$, $|t|\leq 1 $ and 
$\eta\in B_{r_\theta}(\theta)$. 
Then extract a finite complex covering of $[0,\pi]$ made by balls 
$B_{r_{\theta_i}}(\theta)$,
$i= 1,\cdots,M$. Let ${\cal K}''_{p1}$ be the union of the sets of this finite 
covering. The item (1) holds true
for any  $a>0$ with $a\leq b:= Min
\{ b_{\theta_i}\:|\: i= 1,\cdots,M\}$, since
the complex geodesic segment corresponding 
to the metric ${\bf g}_{(\lambda \theta)}$ with $\theta \in {\cal K}''_{p1}$
 from $q\in B_b(\vec{0})$ to $q'\in B_b(\vec{0})$
is the map $t\mapsto exp_{(\theta)}(t {\bf X})$ where $|t|\leq 0$
and $(\theta, {\bf X}) = \Phi^{-1}(\theta, q,q') 
\in B_{r_{\theta_k}}\times
 B_{b_{\theta_k}}(\vec{0})$ for some $k\in \{1,\cdots, M\}$.\\
Let us consider the item (2). Fixed any $\theta\in [0,\pi]$ and 
the positive reals $r_\theta$,
 $b_\theta$ and $V_\theta$ exactly  as those 
in the proof of item (1), choosing $b'_\theta>0$ and $\delta_\theta>0$ 
small enough,
 we can fix an open subset of $B_{r_\theta}(\theta)\times 
V_\theta$,  which is a neighborhood of $(\theta,\vec{0},{\bf 0})$
with the form $B_{r'_\theta}\times 
B_{b'_\theta}(\vec{0}) \times B_{\delta_\theta}$, 
where $0<r'_{\theta}<r_\theta$,
$0<b'< b $ 
and $B_{\delta_\theta}$ is an open ball of radius $\delta_\theta>0$  
 and center in ${\bf 0}\in 
\C^n$. All the tangent spaces $T_q({\cal G})$, $q\in B_{b'}(\vec{0})$,
have been 
identified with $\C^n$ trough the bases induced by the considered
 coordinates. Then, choose an open neighborhood $B_{r''_\theta}(\theta)
\times B_{b''_\theta}(\vec{0})
\times B_{b''}(\vec{0}) \subset \Phi(B_{r'_\theta}(\theta)\times
B_{b'}(\vec{0}) \times B_\delta)$.
Finally notice that, if $q\in B_{b''_\theta}(\vec{0})$ and
$\eta \in B_{r''_\theta}(\theta)$, since $\Phi$ 
is a diffeomorphism in $B_{r'_\theta}(\theta)\times
B_{b'_\theta}(\vec{0}) \times B_{\delta_\theta}$, 
we have 
\beq
\{\eta\} \times \{q\}\times B_{b''_\theta}(\vec{0}) \subset 
\Phi(\{\eta\}\times \{q\}\times B_{\delta_\theta}) \nonumber\:,
\eeq
and in particular, from the definition of $\Phi$,
\beq
B_{b''_\theta}(\vec{0}) \subset exp_{(\eta)q}(B_{\delta_\theta})\:,
\eeq
for any $\eta \in B_{r''_\theta}$. All that can be performed for any fixed
$\theta \in [0,\pi]$. Therefore,
as we done before, we can extract a finite covering of $[0,\pi]$
made by balls $B_{r''_{\theta_k}}(\theta)$, $k=1,\cdots, L$
with union ${\cal K}''_{p2}$. Then put
$b'' := Min\{ b''_{\theta_k}\:|\: k=1,\cdots, L\}$. 
Then $B_{a}(\vec{0})$ is a totally normal neighborhood, if
$a\leq b''$,  with respect to all the metrics ${\bf g}_{(\lambda \theta)}$
whenever $\theta\in {\cal K}''_{p2}$. In ${\cal K}_p'':= {\cal K}''_{p1}
\cap {\cal K}''_{p2}$, (1)  also holds true due to $b''<b$. \\
The proof for the closure of the considered neighborhoods  is trivial
and is obtained by taking $a$ smaller and noticing that 
$\bar{B}_{a'}(\vec{0}) \subset B_{a}(\vec {0})$ if $a'<a$.
$\:\:\:\:\:\:\Box$\\

\noindent
To complete the proof of the item (a), let $0<\rho<a (<c)$ and let $q,q'$ be
any pair of points in $B_\rho(p)$ ($\bar{B}_\rho(p)$), $p\equiv \vec{0}$. 
Let $\vec{u} = \vec{u}_\theta(s)$, $s\in [0,1]$ 
the real-parameter segment geodesic from $q$ to $q'$ in $B_{c}(p)$ (see (p1))
computed with respect to the metric ${\bf g}_{(\lambda\theta)}$ with
$\theta$ arbitrarily fixed in the complex open neighborhood of
$[0,\pi]$ given by ${\cal K}_p:= {\cal K}_p'\cap {\cal K}_p''$. 
We shall show that this real-parameter
segment geodesic lies completely in $B_\rho(p)$ ($\bar{B}_\rho(p)$). 
Consider the function 
$s\mapsto F_\theta(s)$ defined 
in (\ref{Ftheta}) along this geodesic  segment. Assume that 
$F_\theta(s) \geq \rho^2$ 
($F_\theta(s) > \rho^2$ ) 
for some $s_\theta$, that is, $\vec{u}_\theta(s_\theta)$ lies outside 
$B_\rho(p)$ ($\bar{B}_\rho(p)$ )
for some $s_\theta$. Let $s_{\theta0}$, $s_{\theta0}\in ]0,1[$, 
be the value for which $F_\theta$
attains the maximum, say, $\rho_{\theta}^2\geq \rho^{2}$
 ($\rho_\theta^2> \rho$). Then
\beq
0= \frac{dF_\theta}{ds}|_{s=s_{\theta 0}} \:.
\eeq
This means that the real-parameter geodesic segment is tangent to the sphere
$S_{\rho_\theta}(p)$ at the point $\vec{x}(s_{\theta0})$. 
By the choice of $\rho$ the 
considered real-parameter geodesic segment lies inside the sphere
 $S_{\rho_\theta}(p)$, contradicting (p1). \\

\noindent To end the proof, let us prove that in any set ${\cal G}_{pj}$
and  for $\theta \in {\cal K}_p$, the van Vleck-Morette
determinant can be defined as a single-valued function which coincides with
the ordinary van Vleck-Morette determinant for whatever value of 
$\theta$ such that the metric is real, in particular $\theta =0,\pi$.\\
By (\ref{VVMC}), we can assume  that $\Delta_{VVM\theta}^{1/2}$ 
is single-valued,
if the functions defined in our coordinates by 
\beq
(\theta,\vec{x},\vec{y})
 \mapsto  F(\theta,\vec{x},\vec{y})&:=& 
\frac{g_\theta(\vec{x})}{g_\theta(\vec{y})}\nonumber\:,\\
(\theta,\vec{x},\vec{y}) \mapsto G(\theta,\vec{x},\vec{y})&:=& 
\frac{(-1)^D}{ g_\theta(\vec{x})} 
det\left(\frac{\partial^2\sigma_\theta(\vec{x},\vec{y})}{\partial 
x^a\partial y^b}\right) \nonumber\:,
\eeq
take values
away from the cut of a folder  of the domain of definition 
of the functions $z\mapsto z^{1/4}$ and $z\mapsto z^{1/2}$. 
From now on,
 we fix this cut along the negative real axis and work in the folder where
 both  the functions $z\mapsto z^{1/4}$ and $z\mapsto z^{1/2}$
 produce   {\em real and positive} 
values when evaluated on {\em positive real} numbers. 
Let us prove that we can shrink the open set $B_\rho(p)$ 
in (\ref{palla}) (used to define the class of 
${\cal G}_{pj}:= B_{(1+\tanh j) \rho/2}(p)$)
and  ${\cal K}_p$  such that, 
in ${\cal K}_p\times B_\rho(p)\times B_\rho(p)$,  
the functions $F$ and $G$
 take values with strictly positive real part.\\
Fix $\theta \in [0,\pi]$ and $p\equiv\vec{z}$ (the arbitrary center of
$B_\rho(p)$). Trivially  $F(\theta,\vec{z},\vec{z})= 1$.  
On the other hand it also holds 
$\Delta_{VVM\theta}(\vec{z},\vec{z})=1$, because the VVM determinant is a 
bi-scalar and this result can be trivially obtained in normal coordinates 
centered in $\vec{x}$ by (\ref{VVMN}). Therefore we have also, in our 
coordinates, $G(\theta,\vec{z},\vec{z})= 1$.  Since $F$ and $G$ are
jointly continuous in $(\theta,\vec{x},\vec{y})$, there is a neighborhood
of $(\theta,\vec{z},\vec{z})$ of the form $B_{k_{\theta}}(\theta)
\times B_{\rho_\theta}(p)\times B_{\rho_\theta}(p)$, $0<\rho_\theta\leq \rho$, 
where both the functions assume
only 
values with strictly positive real part. This procedure can be performed
 for any point $\theta \in [0,\pi]$, obtaining a covering of this set made of 
complex open disks $B_{k_{\theta}}(\theta)$. 
By compactness, we can extract a finite
sub-covering made by disks $B_{k_{\theta_i}}(\theta_i)$
centered in $\theta_i$, $i=1,\cdots,L$, and a corresponding finite class
of open 
neighborhood of $p$, $B_{\rho_{\theta_i}}(p)$, $i=1,\cdots,L$. 
Then we can take $\rho'>0$
such that $B_{\rho'}(p)
\subset \cap_i B_{\rho_{\theta_i}}(p)$ and use $B_{\rho'}(p)$  
 to define the class of ${\cal G}_{pj}$ by (\ref{palla'}).
Finally ${\cal K}_p$ can be re-defined as the intersection between
the initial ${\cal K}_p$ and $\cup_i B_{k_{\theta_i}}(\theta_i)$.
With these definition both functions $F$ and $G$ assume values with 
strictly positive imaginary part whenever $\theta\in {\cal K}_p$
and $\vec{x},\vec{y}$ belong to any ${\cal G}_{pj}$.
This implies that, in any ${\cal G}_{pj}$, 
$\Delta_{VVM\theta}^{1/2}$ can be defined as a 
single-valued function for any $\theta \in {\cal K}_p$. Moreover, with
the choice above of the folder of definition 
of the functions $z\mapsto z^{1/4}$ and $z\mapsto z^{1/2}$, 
$\Delta_{VVM\theta}^{1/2}(\vec{x},\vec{y})$
 coincides to the usual real one for these $\theta$ such that
 the function takes real values.
$\:\:\:\:\:\:\:\Box$\\

\noindent {\bf Note added.}\\

S. Hollands pointed out to me that a partial but relevant result
about the symmetry of Lorentzian Hadamard coefficients $v_j(x,y)$
in (\ref{V}) is contained in the final chapter of Friedlander's book,
{\em The wave equation on a curved space-time} (Cambridge University Press,
Cambridge, 1975). This result gives an overlap with  results of the present
work in a direct corollary of Theorem 6.4.1 in Fiedlander's book
and taking into account the comment after Theorem 4.3.1 where it is
specified  that the coefficients  considered in the book
essentially coincide with $v_j(x,y)$ Hadamard's coefficients despite
a different use and definition. This corollary and the comment show that,
in a smooth Lorentzian manifold, when $x,y$ belong to a common, sufficiently
small, {\em causal domain} and $\sigma(x,y) < 0$ is satisfied,
then $v_j(x,y) = v_j(y,x)$. This result is achieved using the whole theory
of Lorentzian distribution developed in the book (see in particular
Theorems 5.2.1 and 6.3.2) and makes use of the {\em method of descent}
which explicitly requires a {\em Lorentzian} ($D$ dimensional) metric.
Finally, within a short remark after Theorem 6.4.1, it is suggested that
a generalization to a whole causal domain may be obtained in the analytic case.
Then, it is argued that the smooth non-analytic case also could be enconpassed
by means os somehow approximation procedure of smooth differential equations
by analytic differential equations. This last part of the suggested procedure
seems to be exactly what we explicitly done in {\bf Theorem 3.2}.\\
I am very grateful to S. Hollands for his remark.\\


\begin{thebibliography}{References}






\bibitem[BD82]{bd}
N. D. Birrel and P. C. W.  Davies,
{\em Quantum Fields in Curved Space} (Cambridge University Press,
Cambridge, 1982).

\bibitem[Ca90]{ca} R. Camporesi, Phys. Rep. {\bf 196}, 1 (1990).

\bibitem[Cs96]{cassa} A. Cassa,  Class. Quant. Grav.  {\bf 12}, 5, 1151 (1995).




\bibitem[Ch84]{ch}
I. Chavel,
{\em Eigenvalues in Riemannian Geometry} (Academic Press, Inc., Orlando,
 USA, 1984).


\bibitem[Fu91]{fu}
S.A. Fulling,
 {\em Aspects of Quantum Field Theory in Curved
Space-Time} (Cambridge University Press, Cambridge, 1991).

\bibitem[FR87]{FR} S.A. Fulling and S.N.M. Ruijsenaars, Phys. Rep. {\bf 152},
135 (1987).

\bibitem[FSW78]{fsw} S.A. Fulling, M. Sweeny, R. M. Wald,
Comm.. Math. Phys. {\bf 63}, 257 (1978)

\bibitem[Ga64]{garabedian}
P. R. Garabedian, {\em Partial Differential Equations}
(John Wiley and Sons, Inc., New York, 1964).


\bibitem[GH93]{EQG} {\em Euclidean Quantum Gravity} Editors
G.W. Gibbons, S.W. Hawking (World Scientific, Singapore, 1993).  


\bibitem[Gi84]{gi}
P.G. Gilkey, {\em Invariance theory, the heat equation and the
Atiyah-Singer index theorem} Math. lecture series 11 (Publish or
Perish Inc. Boston, Ma., 1984)


\bibitem[Ha77]{ha}
S. W. Hawking, Comm.. Math. Phys. {\bf 55}, 133 (1977).

\bibitem[KN63]{kn} S. Kobayashi and K. Nomizu,
{\em Foundations of Differential Geometry} Vol.1
(Interscience Publishers, New York, 1963)


\bibitem[LB83]{lb} C. LeBrun, Trans. A.M.S. {\bf 278}, 1, 209 (1983). 


\bibitem[Mo99a]{m1} V. Moretti,  Comm.. Math. Phys. 201, 327 (1999).

\bibitem[Mo99b]{m2} V. Moretti, J. Math. Phys. 40, 3843 (1999). 

\bibitem[Mo99c]{m3} V. Moretti,  Comm. Math. Phys. 208, 283 (1999).


\bibitem[ON83]{oneill} B. O'Neill, {\em Semi-Riemannian Geometry with 
applications to Relativity.} (Academic Press, New York, (1983)).



\bibitem[Wa78]{wald78}
R. M. Wald, Phys. Rev. D {\bf 17}, 1477 (1978).

\bibitem[Wa79]{wald79}
R. M. Wald, Comm.. Math. Phys. {\bf 70},  226 (1979).

\bibitem[Wa84]{waldR}  R.M. Wald,
{\em General Relativity} (The University of Chicago Press, Chicago,
1984).

\bibitem[Wa94]{wald94}  R.M. Wald,
{\em Quantum Field theory and Black Hole
Thermodynamics
in Curved Spacetime} (The University of Chicago Press, Chicago,
1994).


\end{thebibliography}
 \end{document}